\documentclass[article,draftcls,onecolumn]{IEEEtran}
\usepackage{amsmath,amssymb}
\usepackage{graphicx,color}
\usepackage{subfig}
\usepackage{cite}
\usepackage{epstopdf}
\usepackage{url,doi}
\usepackage{caption}
\usepackage[colorinlistoftodos,color=orange!70,textsize=scriptsize,shadow]{todonotes}
\usepackage{setspace}

\usepackage[normalem]{ulem}
\newcommand\redout{\bgroup\markoverwith{\textcolor{red}{\rule[.5ex]{2pt}{0.4pt}}}\ULon}

\newtheorem{theorem}{Theorem}

\newtheorem{corollary}{Corollary}

\newtheorem{lemma}{Lemma}

\newtheorem{assumption}{\textbf{Modelling Assumption}}

\def\mbf{\mathbf}

\def\msf{\mathsf}
\newcommand{\hp}{\mathsf{HP}}
\newcommand{\lp}{\mathsf{LP}}
\newcommand{\normal}{\mathsf{Normal}}
\newcommand{\alerted}{\mathsf{Alerted}}

\newcommand{\consumer}{consumer}
\newcommand{\lambdana}{\lambda_{N,A}}
\newcommand{\lambdaaa}{\lambda_{A,A}}
\newcommand{\lambdapna}{\lambda'_{N,A}}
\newcommand{\lambdapaa}{\lambda'_{A,A}}

\newcommand{\rhn}{C_{HN}}
\newcommand{\rha}{C_{HA}}
\newcommand{\rhaone}{C_{HA_1}}
\newcommand{\rhak}{C_{HA_K}}
\newcommand{\cl}{C_{L}}

\newcommand{\E}{\mathbb{E}}
\newcommand{\argmax}{\mathop{\mathrm{argmax}}}

\newcommand{\unif}{\mathsf{Unif}}
\begin{document}

\title{Designing Incentive Schemes For Privacy-Sensitive Users}
\author{
Chong Huang, 
\and Lalitha Sankar 
\and and Anand~D.~Sarwate 
\thanks{C. Huang and L. Sankar are with the Department of Electrical, Computer, and Energy Engineering at Arizona State University, Tempe, AZ 85287 (e-mail: \texttt{chong.huang@asu.edu}, \texttt{lalithasankar@asu.edu}).  A.D. Sarwate is with the Department of Electrical and Computer Engineering at Rutgers, the State University of New Jersey, Piscataway, NJ 08854 (e-mail: \texttt{asarwate@ece.rutgers.edu}).}%
}

\maketitle

\pagenumbering{arabic}

\begin{abstract}
Businesses (\textit{retailers}) often wish to offer personalized advertisements (\textit{coupons}) to individuals (\textit{consumers}), but run the risk of strong reactions from consumers who want a customized shopping experience but feel their privacy has been violated. Existing models for privacy such as differential privacy or information theory try to \textit{quantify} privacy risk but do not capture the subjective experience and heterogeneous \textit{expression} of privacy-sensitivity. We propose a Markov decision process (MDP) model to capture (i) different consumer privacy sensitivities via a time-varying state; (ii) different coupon types (action set) for the retailer; and (iii) the action-and-state-dependent cost for perceived privacy violations. For the simple case with two states (``Normal" and ``Alerted"), two coupons (targeted and untargeted) model, and consumer behavior statistics known to the retailer, we show that a stationary threshold-based policy is the optimal coupon-offering strategy for a retailer that wishes to minimize its expected discounted cost. The threshold is a function of all model parameters; the retailer offers a targeted coupon if their belief that the consumer is in the "Alerted" state is below the threshold. We extend this two-state model to consumers with multiple privacy-sensitivity states as well as coupon-dependent state transition probabilities. Furthermore, we study the case with imperfect (noisy) cost feedback from consumers and uncertain initial belief state. 

\textbf{Keywords-}Privacy, Markov decision processes, retailer-consumer interaction, optimal policies.

\end{abstract}

\section{Introduction}
Programs such as retailer ``loyalty cards'' allow companies to automatically track a customer's financial transactions, purchasing behavior, and preferences. They can then use this information to offer customized incentives, such as discounts on related goods. Consumers may benefit from retailer's knowledge by using more of these targeted discounts or coupons while shopping. However, in some cases the coupon offer implies that the retailer has learned something sensitive or private about the consumer.  For example, a retailer could infer a \consumer's pregnancy~\cite{hill2012target}. Such violations may make consumers skittish about purchasing from such retailers.

However, modeling the privacy-sensitivity of a consumer is not always straightforward: widely-studied models for quantifying privacy risk using differential privacy or information theory do not capture the subjective experience and heterogeneous \textit{expression} of consumer privacy. The goal of this paper is to introduce a framework to model the consumer-retailer interaction problem and better understand how retailers can develop coupon-offering policies that balances their revenue objectives while being sensitive to consumer privacy concerns. The main challenge for the retailer is that the consumer's responses to coupons are not known \textit{a priori}; furthermore, consumers do not ``add noise" to their purchasing behavior as a mechanism to stay private. Rather, the offer of a coupon may provoke a reaction from the consumer, ranging from ``unaffected" to ``ambiguous" or ``partially concerned" to ``creeped out." This reaction is mediated by the consumer's sensitivity level to privacy violations, and it is these levels that we seek to model via a Markov decision process. These privacy-sensitivity states of the consumers are often revealed to the retailer through their purchasing patterns. In the simplest case, they may accept or reject a targeted coupon. We capture these aspects in our model and summarize our main contributions below.
\subsection{Main Contributions} 
We propose a partially-observed Markov decision process (POMDP) model for this problem in which the consumer's state encodes their privacy sensitivity, and the retailer can offer different levels of privacy-violating coupons. The simplest instance of our model is one with two states for the consumer, denoted as ``Normal" and ``Alerted," and two types of coupons: untargeted \textit{low privacy} (LP) or targeted \textit{high privacy} (HP).  At each time, the retailer may offer a coupon and the consumer transitions from one state to another according to a Markov chain that is independent of the offered coupon.  The retailer suffers a cost that depends both on the type of coupon offered and the state of the \consumer. The costs reflect the advantage of offering targeted HP coupons relative to untargeted LP ones while simultaneously capturing the risk of doing so when the consumer is already ``Alerted". 

Under the assumption that the retailer (via surveys or prior knowledge) knows the statistics of the consumer Markov process, i.e., the likelihoods of becoming ``Alerted" and staying ``Alerted", and a belief about the initial consumer state, we study the problem of determining the optimal coupon-offering policy that the retailer should adopt to minimize the long-term discounted costs of offering coupons. We extend the simple model above to multiple states and coupon-dependent transitions. We model the latter via two Markov processes for the consumer, one for each type (HP or LP) of coupon such that a persnickety consumer who is easily ``Alerted" will be more likely to do so when offered an HP (relative to LP) coupon. Furthermore, for noisy costs, we propose a heuristic method to compute the decision policy. Moreover, if the initial belief state is unknown to the retailer, we use a Bayesian model to estimate the belief state. Our main results can be summarized as follows:
\begin{enumerate}
\item  There exists an optimal, stationary, threshold-based policy for offering coupons such that a HP coupon is offered only if the belief of being in the ``Alerted" state at each interaction time is below a certain threshold; this threshold is a function of all the model parameters. This structural result holds for multiple states and coupon-dependent transitions.
\item The threshold for offering a targeted HP coupon increases in the following cases:
\begin{enumerate}
\item once ``Alerted,'' the consumer remains so for a while -- the retailer is more willing to take risks since the the consumer takes a while to transition to ``Normal";
\item the consumer is very unlikely to get ``Alerted"; 
\item the cost of offering an untargeted LP coupon is high and close to the cost of offering a targeted HP coupon to an ``Alerted" consumer; and 
\item when the retailer does not discount the future heavily, i.e., the retailer stands to benefit by offering HP coupons for a larger set of beliefs about the consumer's state.
\end{enumerate}
\item For the coupon-dependent Markov model for the consumer, the threshold is smaller than for the non-coupon dependent case which encapsulates the fact that highly sensitive consumers will force the retailers to behave more conservatively.
\item By adopting a heuristic threshold policy computed by the mean value of costs, the retailer can minimize the discounted cost effectively even if costs are noisy. Moreover, the Bayesian approach helps the retailer to estimate the \consumer\ state when the initial belief state is unknown. 
\end{enumerate}
Our results use many fundamental tools and techniques from the theory of MDPs through appropriate and meaningful problem modeling. We briefly review the related literature in consumer privacy studies as well as MDPs.

\subsection{Related Work}Several economic studies have examined consumer's attitudes towards privacy via surveys and data analysis including studies on the benefits and costs of using private data (e.g., Aquisti and Grossklags in~\cite{acquisti2010economics}). On the other hand, formal methods such as differential privacy are finding use in modeling the value of private data for market design~\cite{ghosh2013selling} and for the problem of partitioning goods with private valuation function amongst the agents~\cite{hsu2013private}. In these models the goal is to elicit private information from individuals. Venkitasubramaniam~\cite{venkitasubramaniam2013privacy} recently used an MDP model to study data sharing in control systems with time-varying state. He minimizes the weighted sum of the utility (benefit) that the system achieves by sharing data (e.g., with a data collector) and the resulting privacy leakage, quantified using the information-theoretic equivocation function. In our work we do not quantify \textit{privacy loss} directly; instead we model \textit{privacy-sensitivity} and resulting user behavior via MDPs to determine interaction policies that can benefit both consumers and retailers. 
To the best of our knowledge, a formal model for \consumer-retailer interactions and the related privacy issues has not been studied before; in particular, our work focuses on explicitly considering the consequence to the retailer of the \consumer s' awareness of privacy violations.

Markov decision processes (MDPs) have been widely used for decades across many fields~\cite{feinberg2002handbook,puterman2009markov}; in particular, our model is related to problems in control with communication constraints~\cite{lipsa2011remote,nayyar2013optimal} where state estimation has a cost. Our costs are action and state dependent and we consider a different optimization problem. Classical target-search problems~\cite{macphee1995optimal} also have optimal policies that are thresholds, but in our model the retailer goal is not to estimate the consumer state but to minimize cost. The model we use is most similar to Ross's model of product quality control with deterioration~\cite{ross1971quality}, which was more recently used by Laourine and Tong to study the Gilbert-Elliot channel in wireless communications~\cite{laourine2010betting}, in which the channel has two states and the transmitter has two actions (transmit or not). We cannot apply their results directly due to our different cost structure, but use ideas from their proofs. Furthermore, we go beyond these works to study privacy-utility tradeoffs in consumer-retailer interactions with more than two states and action-dependent transition probabilities. We apply more general MDP analysis tools to address our formal behavioral model for privacy-sensitive consumers.

While the MDP model used in this paper is simple, its application to the problem of revenue maximization with privacy-sensitive consumers is novel. We show that the optimal stationary policy exists and it is a threshold on the probability of the consumer being alerted. We extend the model to cases of \consumer s with multiple states and \consumer s with coupon-dependent transition probabilities.  Our basic model assumes the probability of the consumer being alerted can be inferred from the received costs. When the costs are stochastic, we use a Bayesian estimator to track this probability and propose a heuristic coupon offering policy for this setting. In the conclusion we describe several other interesting avenues for future work.

The paper is organized as follows: Section \ref{sec:sysmod} introduces the system model and its extensions. The main result for known \consumer\ statistics are presented in Section \ref{sec:optpolicy}. Section \ref{sec:cpdep} and \ref{sec:bayes} discuss optimal stationary policy results for \consumer s with coupon dependent response and noisy costs with unknown initial belief, respectively. Finally, some concluding remarks and future work are provided in Section \ref{sec:conclusion}. 


\section{System Model}
\label{sec:sysmod}
We model interactions between a retailer and a \consumer\ via a discrete-time system (Figure~\ref{fig:figure1}). At each time $t$, the \consumer\ has a discrete-valued state and the retailer may offer one of two coupons: high privacy risk~($\hp$) or low privacy risk~($\lp$). 
The \consumer\ responds to the personalized coupon by imposing a cost on the retailer that depends on the coupon offered and its own state. For example, a \consumer\ who is ``alerted'' (privacy-aware) may respond to an $\hp$ coupon by imposing a high cost to the retialer, such as reducing purchases at the retailer. The retailer's goal is to decide which type of coupon to offer at each time $t$ to minimize its cost.


\subsection{Consumer with Two States and Coupon Independent Transitions.\label{sec:model:ext:basicmodel}} 
\subsubsection{Consumer Model} 
\begin{assumption}{\textbf{(Consumer's state)}}
\textit{We model the \consumer's response to coupons by assuming them to be in one of several states. Each state corresponds to a type of \consumer\ behavior in terms of purchasing (Privacy sensitivity).}
\end{assumption}

\begin{figure}
\centering
\includegraphics[width=70mm]{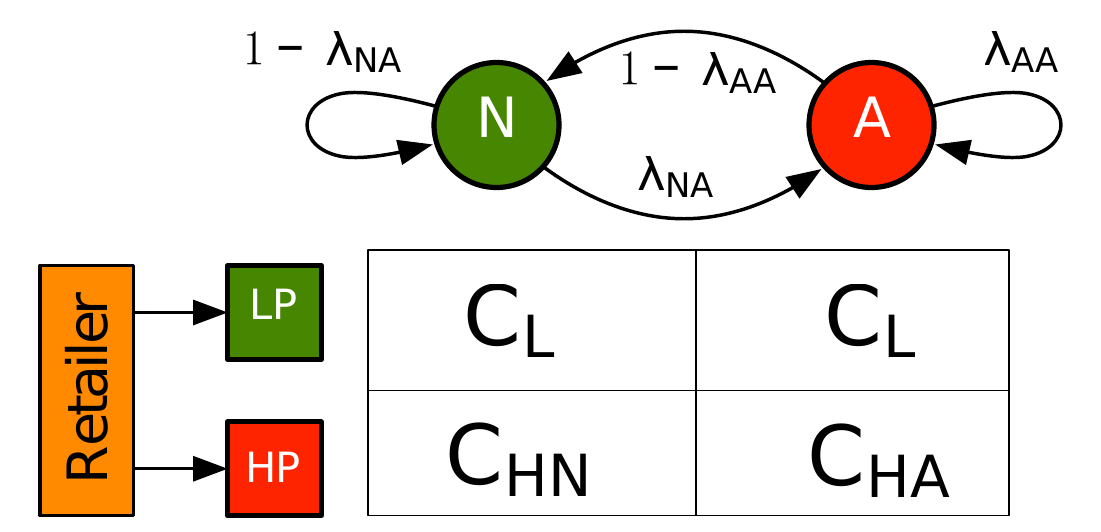}
\caption{Markov state transition model for a two-state consumer.}
\label{fig:figure1}
\end{figure}

For this paper, we first focus on the two-state case; the consumer may be $\normal$ or $\alerted$. Later we will extend this model to multiple \consumer\ states, \consumer\ with coupon dependent response, and unknown initial \consumer\ state cases. The \consumer\ state at timet $t$ is denoted by $G_t\in\{\normal,\alerted\}$. If a \consumer\ is in $\normal$ state, the \consumer\ is less sensitive to coupons from the retailer in terms of privacy. However, in the $\alerted$ state, the \consumer\ is likely to be more sensitive to coupons offered by the retailer, since it is more cautious about revealing information to the retailer. The evolution of the \consumer\ state is modeled as a infinite-horizon discrete time Markov chain (Figure~\ref{fig:figure1}). The \consumer\ starts out in a random initial state unknown to the retailer and the transition of the \consumer\ state is independent of the action of the retailer. A \emph{belief state} is a probability distribution over possible states in which the \consumer\ could be. The belief of the \consumer\ being in $\alerted$ state at time $t$ is denoted by $p_t$. We define $\lambdana=Pr[G_t=\alerted|G_{t-1}=\normal]$ to be the transition probability from $\normal$ state to $\alerted$ state and $\lambdaaa=Pr[G_t=\alerted|G_{t-1}=\alerted]$ to be the probability of staying in $\alerted$ state when the previous state is also $\alerted$. The transition matrix $\mathbf{\Lambda}$ of the Markov chain can be written as
\begin{equation}
\label{eq:transition}
\mathbf{\Lambda} =
\begin{pmatrix}
   1-\lambdana & \lambdana \\
   1-\lambdaaa & \lambdaaa 
  \end{pmatrix}.
\end{equation}
We assume the transition probabilities are known to the retailer; this may come from statistical analysis such as a survey of \consumer\ attitudes. The one step transition function, defined by 
	\begin{align}
	T(p_t)=(1-p_t)\lambdana+ p_t\lambdaaa, 
	\label{eq:T}
	\end{align}
represents the belief that the \consumer\ is in $\alerted$ state at time $t+1$ given $p_t$, the $\alerted$ state belief at time $t$.

\begin{assumption}{\textbf{(State transitions)}}
\label{consumertransition}
\textit{Consumers have an inertia in that they tend to stay in the same state. Moreover, once \consumer s feel their privacy is violated, it will take some time for them to come back to $\normal$ state. }
\end{assumption}

The above assumption implies $\lambdaaa \ge 1-\lambdaaa$, $1-\lambdana \ge \lambdana$, and $\lambdana \ge 1-\lambdaaa$. Thus, by combining the above three inequalities, we have $\lambdaaa \ge \lambdana$.


\subsubsection{Retailer Model} 

At each time $t$, the retailer can take an \textit{action} by offering a coupon to the \consumer. We define the action at time $t$ to be $u_t\in \{\hp ,\lp\}$, where $\hp$ denotes offering a high privacy risk coupon (e.g. a targeted coupon) and $\lp$ denotes offering a low privacy risk coupon (e.g. a generic coupon). The retailer's utility is modeled by a \emph{cost} (negative revenue) which depends on the \consumer's state and the type of coupon being offered. If the retailer offers an $\lp$ coupon, it suffers a cost $\cl$ independent of the consumer's state: offering $\lp$ coupons does not reveal anything about the state.  However, if the retailer offers an $\hp$ coupon, then the cost is $\rhn$ or $\rha$ depending on whether the \consumer's state is $\normal$ or $\alerted$. Offering an $\hp$ (high privacy risk, targeted) coupon to a $\normal$ \consumer\ should incur a low cost (high reward), but offering an $\hp$ coupon to an $\alerted$ consumer should incur a high cost (low reward) since an $\alerted$ \consumer\ is privacy-sensitive. Thus, we assume $\rhn \le \cl \le \rha$.

Under these conditions, the retailer's objective is to choose $u_t$ at each time $t$ to minimize the total cost inccured over the entire time horizon. The $\hp$ coupon reveals information about the state through the cost, but is risky if the \consumer\ is alerted, creating a tension between cost minimization and acquiring state information.

\subsubsection{Minimum Cost Function}

We define $C(p_t,u_t)$ to be the expected cost acquired from an individual \consumer\ at time $t$ where $p_t$ is the probability that the consumer is in Alerted state and
$u_t$ is the retailer's action: 
\begin{align}
\label{instcost}
C(p_t,u_t)=
\left\{
\begin{array}{ll}
\cl  & \mbox{if } u_t=\lp \\
(1-p_t)\rhn+p_t\rha & \mbox{if } u_t=\hp
\end{array}
\right..
\end{align}
Since the retailer knows the \consumer\ state from the incurred cost only when an $\hp$ coupon is offered, the state of the \consumer\ may not be directly observable to the retailer. Therefore, the problem is actually a Partially Observable Markov Decision Process (POMDP)~\cite{bertsekas1995dynamic}.
  
We model the cost of violating a \consumer's privacy as a short term effect. Thus, we adopt a discounted cost model with discount factor $\beta\in(0,1)$.
At each time $t$, the retailer has to choose which action $u_t$ to take in order to minimize the expected discounted cost over infinite horizon. A policy $\pi$ for the retailer is a rule that selects a coupon to offer at each time. Thus, given that the belief of the \consumer\ being in $\alerted$ state at time $t$ is $p_t$ and the policy is $\pi$, the infinite-horizon discounted cost starting from $t$ is
\begin{equation}
\label{policy}
V^{\pi,t}_\beta(p_t)=\mathbb{E}_\pi\left[\sum\limits_{i=t}^{\infty}\beta^iC(p_i,u_i)|p_t\right],
\end{equation}
where $\mathbb{E}_\pi$ indicates the expectation over the policy $\pi$.
The objective of the retailer is equivalent to minimizing the discounted cost over all possible policies. Thus, we define the minimum cost function starting from time $t$ over all policies to be 
\begin{equation}
\label{minpolicy}
V^t_\beta(p_t)=\min\limits_{\pi}{V^{\pi,t}_\beta(p_t)}\text{      for all }p_t\in [0,1].
\end{equation}

We define $p_{t+1}$ to be the belief of the \consumer\ being in $\alerted$ state at time $t+1$. The minimum cost function $V^t_\beta(p_t)$ satisfies the Bellman equation~\cite{bertsekas1995dynamic}:
\begin{equation}
\label{Bellman1}
V^t_\beta(p_t)=\min\limits_{u_t \in \{\hp,\lp\}}\{V^t_{\beta,u_t}(p_t)\},
\end{equation}
\begin{equation}
\label{Bellman2}
V^t_{\beta,u_t}(p_t)=\beta^tC(p_t,u_t)+V^{t+1}_\beta(p_{t+1}|p_t,u_t).
\end{equation}

An optimal policy is \textit{stationary} if it is a deterministic function of states, i.e., the optimal action at a particular state is the optimal action in this state at all times. We define $\mathcal{P}=\{[0,1]\}$ to be the belief space and $\mathcal{U}=\{\lp,\hp\}$ to be the action space. In the context of our model, the optimal stationary policy is a deterministic function mapping $\mathcal{P}$ into $\mathcal{U}$. Since the problem is an infinite-horizon, finite state and finite action MDP with discounted cost, by~\cite{ross2013applied}, there exists an optimal stationary policy $\pi^*$ such that starting from time $t$,
\begin{equation}
V^t_\beta(p_t)=V_{\beta}^{\pi^*,t}(p_t).
\end{equation}
Thus, only the optimal stationary policy is considered because it is tractable and achieves the same minimum cost as any optimal non-stationary policy.
 
By~(\ref{Bellman1}) and~(\ref{Bellman2}), the minimum cost function evolves as follows. If an $\hp$ coupon is offered at time $t$, the retailer can perfectly infer the \consumer\ state based on the incurred cost. Therefore,
\begin{align}
\label{eq:hpevo}
V^t_{\beta, \hp}(p_t) = \beta^tC(p_t,\hp)+(1-p_t)V^{t+1}_\beta(\lambdana) + p_tV^{t+1}_\beta(\lambdaaa).
\end{align}
If an $\lp$ coupon is offered at time $t$, the retailer cannot infer the \consumer\ state from the cost since both $\normal$ and $\alerted$ consumer impose the same cost $\cl$. Hence, the discounted cost function can be written as 
\begin{align}
\label{eq:lpevo}
V^t_{\beta, \lp}(p_t) 
&= \beta^tC(p_t,\lp)+V^{t+1}_\beta(p_{t+1}) \nonumber \\
&= \beta^t\cl+V^{t+1}_\beta(T(p_t)).
\end{align}
Correspondingly, the minimum cost function is given by
\begin{equation}
\label{eq:cphplp}
V^t_\beta(p_t)=\min\{V^t_{\beta, \lp}(p_t),V^t_{\beta, \hp}(p_t)\}.
\end{equation}
We now describe some simple extensions of this basic model.

\subsection{Consumer with Multi-Level Alerted States \label{sec:model:ext:multilevel} }

In this section, the case that the \consumer\ has multiple $\alerted$ states is studied. Without loss of generality, we define $G_t\in\{\normal,\alerted _1,\dots \alerted_K\}$ to be the \consumer\ state at time $t$. If the \consumer s is in $\alerted_k$ state, it is even more cautious about coupons than in $\alerted_{k-1}$ state. Beliefs of the \consumer\ being in $\normal$, $\alerted_1, \dots, \alerted_K$ state at time $t$ are defined by $\bar{\mathbf{p}}_t=(p_{N,t},p_{A_1,t}, \dots, p_{A_K,t})^T$. At each time $t$, the retailer can offer either an $\hp$ or an $\lp$ coupon. Costs of the retailer when an $\hp$ coupon is offered while the state of the \consumer\ is $\normal$, $\alerted_1, \dots, \alerted_K$ are defined by $\bar{\mathbf{C}}=(\rhn,\rhaone, \dots, \rhak)^T$. If an $\lp$ coupon is offered, no matter in which state, the retailer gets a cost of $\cl$. We assume that $\rhak\ge \dots \ge \rhaone\ge \cl\ge \rhn$. The minimum cost function evolves as follows: 
\begin{align}
V^t_\beta(\bar{\mbf{p}}_t)=\min\{V^t_{\beta, \lp}(\bar{\mbf{p}}_t),V^t_{\beta, \hp}(\bar{\mbf{p}}_t)\},
\end{align}
where $V^t_{\beta, \lp}(\bar{\mbf{p}}_t)=\beta^t\cl+V^{t+1}_\beta(\bar{\mbf{p}}_{t+1})
$ and $V^t_{\beta, \hp}(\bar{\mbf{p}}_t)=\beta^t\bar{\mbf{p}}_t^{T}\bar{\mbf{C}}+V^{t+1}_\beta(\bar{\mbf{p}}_{t+1})$ represents the cost of offering an $\lp$ and an $\hp$ coupon, respectively. This model can be generalized to \consumer\  with finitely many states.

\subsection{Consumer with Coupon Dependent Transitions\label{sec:model:ext:cdp} }

\begin{figure}
\centering
\includegraphics[width=80mm]{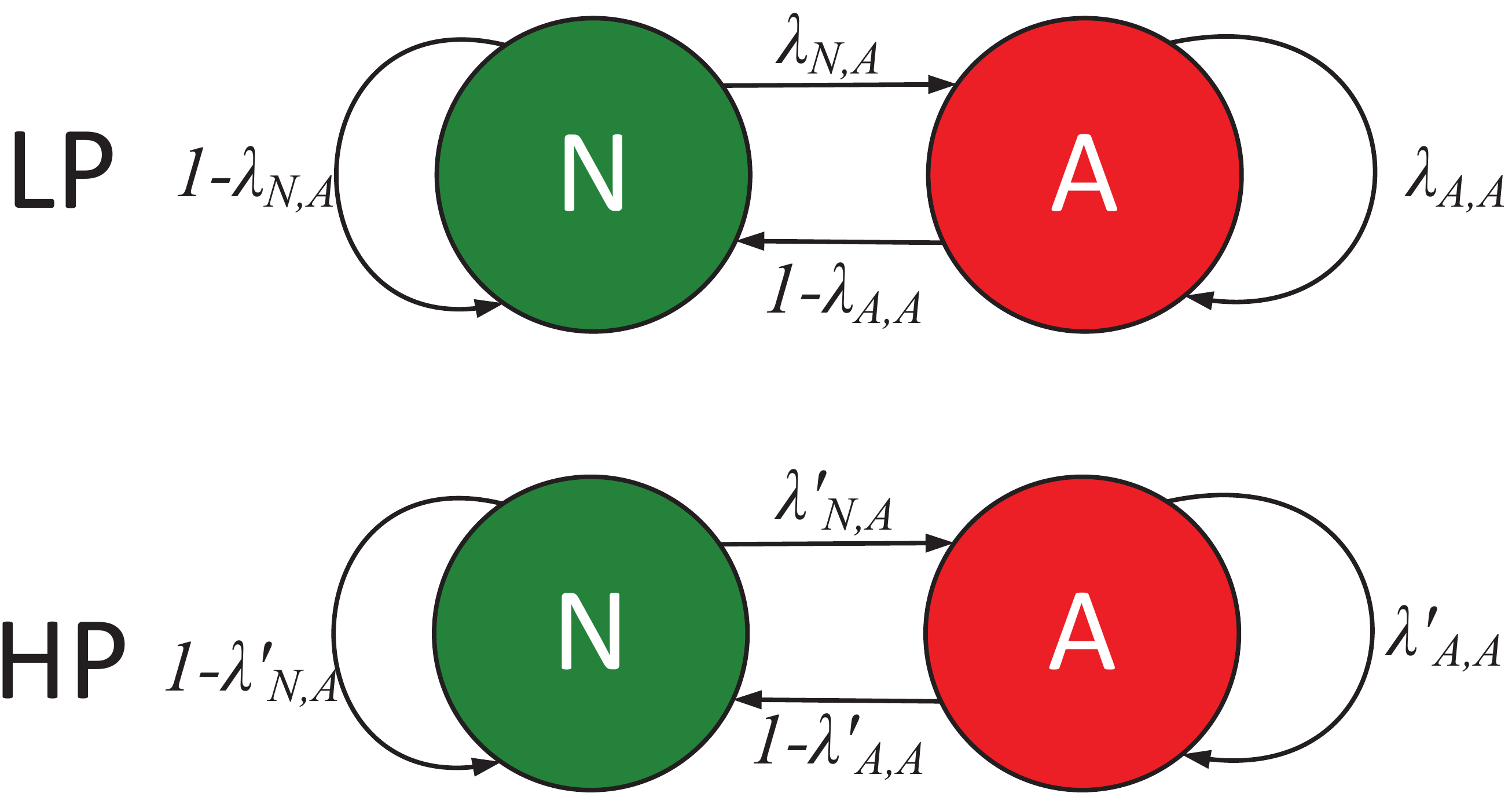}
\caption{Coupon type dependent Markov state transition model.}
\label{fig:figure3}
\end{figure}
In the previous formulations, we assume that the \consumer's state transition is independent of the retailer's action. A natural extension is the case where the action of the retailer can affect the dynamics of the \consumer\ state evolution (Figure \ref{fig:figure3}). Generally, a \consumer's reactions to $\hp$ and $\lp$ coupons are different. For example, a consumer is likely to feel less comfortable when being offered a coupon on medication ($\hp$) than food ($\lp$). Thus, in Section \ref{sec:cpdep}, we assume that the Markov transition probabilities are dependent on the coupon offered with transition matrix given by  $\mbf{\Lambda}_{\msf{LP}}$($\mbf{\Lambda}_{\msf{HP}}$), 
where $\mbf{\Lambda}_{\msf{LP}}$ and $\mbf{\Lambda}_{\msf{HP}}$ are defined as: 
\begin{align}
\mbf{\Lambda}_{\lp} =
  \begin{pmatrix}
  1-\lambdana & \lambdana \\
  1-\lambdaaa & \lambdaaa 
 \end{pmatrix},
 \mbf{\Lambda}_{\hp} =
  \begin{pmatrix}
   1-\lambdapna  & \lambdapna \\
   1-\lambdapaa & \lambdapaa 
  \end{pmatrix}.
  \end{align}
Thus, the minimum cost function is given by \eqref{eq:cphplp}, where 
$V^t_{\beta, \lp}(p_t)=\beta^tC(p_t,\lp)+V^{t+1}_\beta(T(p_t))$ and $V^t_{\beta, \hp}(p_t)=\beta^tC(p_t,\hp)+(1-p_t)V^{t+1}_\beta(\lambdana')+p_tV^{t+1}_\beta(\lambdaaa')$ denotes the cost function of using an $\lp$ coupon and an $\hp$ coupon, respectively. $T(p_t)$ and $T'(p_t)$ are the one step transition given by $T(p_t)=\lambdana(1-p_t)+\lambdaaa p_t$ and $T'(p_t)=\lambdapna(1-p_t)+\lambdapaa p_t$.

\subsection{Policies under Noisy Cost Feedback and Uncertain Initial Belief}

Consider a setting in which the feedback regarding the cost may be noisy, e.g., the cost incurred by the \consumer's response to the coupon is not deterministic. For each individual \consumer, the state transition is independent of the action of the retailer. For given state $G_t$ and action $u_t$, define the distribution of observing a cost $C_t=c$ to be $f(c|G_t,u_t)$. In this case, the threshold policy computed using costs might not be optimal. Moreover, if the initial belief is unknown to the retailer, it has to estimate the \consumer\ state before making decision. Thus, we propose some alternative approaches to decide which coupon to offer when those costs are random. A heuristic approach to deal with the randomized cost is to use the threshold $\tau$ computed by the mean value of costs. Furthermore, the estimation of \consumer\ belief state $p_t$ or the actual state $G_t$ is updated by the maximum a posteriori rule (MAP)~\cite{gelman2014bayesian}. After the estimation process, the retailer decides which coupon to offer based on the threshold policy given in Section \ref{sec:optpolicy}.

\subsection{Summary of Main Results} 
For the problems described in Subsection~\ref{sec:model:ext:basicmodel}, \ref{sec:model:ext:multilevel}, and~\ref{sec:model:ext:cdp}, given all system parameters, we show the following:
\begin{itemize}
\item there exists an optimal stationary solution which has a single threshold property and
\item the threshold only depends on the system parameters, i.e., transition probabilities and instantaneous cost associated with each type of coupon.
\end{itemize} This means by adopting the optimal policy, the retailer will offer an $\hp$ coupon if $p_t$ is less than some threshold and offer an $\lp$ if $p_t$ is above the threshold. 

For the model described in Subsection~\ref{sec:model:ext:cdp}, we assume that the cost feedbacks are noisy and \consumer\ belief state is unknown to the retailer. For this model:
\begin{itemize}
\item we design a heuristic threshold policy when the received costs are noisy.
\item a Bayesian estimation approach is proposed to estimate the actual state or the belief state of the \consumer\ when the initial state is unknown to the retailer.
\end{itemize}      


\section{Optimal Policies with Known Consumer Statistics}
\label{sec:optpolicy}

In this section, we consider  the basic formulation as well as the first three extensions. 
First, we assume that there are only one retailer and one \consumer\ in the system and the state transition of the \consumer\ is independent of the coupon offered. The evolution of the minimum cost function is given in \eqref{eq:hpevo}, \eqref{eq:lpevo}, and \eqref{eq:cphplp}.
\subsection{Properties of Minimum Cost Function}

\begin{lemma}
\label{lm1}
Assume $V^{t,m}_{\beta}$ to be the minimum cost when the decision horizon starts from $t$ and only spans $m$ stages, given a time invariant action set $u_i\in\mathcal{U}=\{\lp,\hp\}$, for any $i=0,1,\ldots$ , $V^{t,m}_\beta(p)=\beta V^{t-1,m}_\beta(p)$. 
\end{lemma}

\begin{IEEEproof}
By \eqref{minpolicy} and $u_i\in\{\lp,\hp\}$ for any $i=0,1,\ldots$.
\begin{align}
\begin{split}
V_{\beta}^{t,m}(p) 
	&=\min\limits_{\pi}\E_{\pi} \left[ \sum\limits_{i=t}^{t+m-1}\beta^iC(p_i,u_i)|p_t=p \right]\\
&=\beta\min\limits_{\pi}\E_{\pi}\left[\sum\limits_{i=t-1}^{t+m-2}\beta^iC(p_i,u_i)|p_{t-1}=p\right]\\
&=\beta V^{t-1,m}_\beta(p).
\end{split}
\end{align}
By using induction on $t$, we can easily prove $V^{t,m}_\beta(p)=\beta V^{t-1,m}_\beta(p)=\cdots=\beta^tV^{0,m}_\beta(p)$.
\end{IEEEproof}
\begin{lemma} 
\label{lm2}
The minimum cost function $V^t_\beta (p)$ is a concave and non-decreasing function of $p$.
\end{lemma}

\begin{IEEEproof}
We prove these properties by induction. Define $V^{t,m}_{\beta}$ to be the minimum cost when the decision horizon starts from $t$ and only spans $k$ stages. For $k=1$,
\begin{align}
\label{k1tcost}
V^{t,k}_\beta(p)=\min\{\cl,(1-p)\rhn+p\rha\},
\end{align}
which is a concave function of $p$. For $k=n-1$, assume that $V^{t,k}_\beta(p)$ is a concave function. Then, for $k=n$, since $V^{t,n-1}_\beta(p)$ is concave and $V^{t,k}_{\beta,\lp}(p) =\beta^t\cl+ V^{t+1,n-1}_{\beta}(T(p))$, by the definition of concavity and Lemma \ref{lm1}, we can conclude that $V^{t,k}_{\beta,\lp}(p)$ is concave. Also,  $V^{t,k}_{\beta,\hp}(p)$ is an affine function of $p$, thus $V^{t,k}_{\beta}(p)=\min\{V^{t,k}_{\beta,\lp}(p),V^{t,k}_{\beta,\hp}(p)\}$ is a concave function of $p$. Taking $k\rightarrow \infty, V^{t,k}_\beta(p)\rightarrow V^t_\beta(p)$,
which implies $V^t_\beta(p)$ is a concave function.

Next, we prove the non-decreasing property of the minimum cost function. For $k=1$, as shown in equation \eqref{k1tcost}, it is a non-decreasing function of $p$. Assume that $V^{t,k}_\beta(p)$ is a non-decreasing function for $k=n-1$. For $k=n$, Let $p_{1}\ge p_{2}$, 
\begin{align} 
& V^{t,k}_{\beta,\lp}(p_{1})-V^{t,k}_{\beta,\lp}(p_{2})\\&  = \beta (V^{t,n-1}_{\beta}(T(p_{1}))-V^{n-1}_{\beta}(T(p_{2})))\\
& = \beta (V^{t,n-1}_{\beta}((\lambdaaa-\lambdana)p_{1}+\lambdana)\nonumber\\& \qquad-V^{t,n-1}_{\beta}((\lambdaaa-\lambdana)p_{2}+\lambdana)))\\
& \ge 0.
\end{align}
By using the same technique, we can prove that given $p_{2}-p_{1}\le 0,\rhn-\rha\le 0 \text{ and }V^{t,k-1}_{\beta}(\lambdana)-V^{t,k-1}_{\beta}(\lambdaaa)\le 0$,
\begin{align}
V^{t,k}_{\beta,\hp}(p_{1})-V^{t,k}_{\beta,\hp}(p_{2})\ge 0.
\end{align} 
Since $V^{t,k}_\beta(p_t)=\min\{V^{t,k}_{\beta,\lp}(p),V^{t,k}_{\beta,\hp}(p)\}$, it is the minimum of two non-decreasing functions. Therefore, $V^{t,k}_\beta(p)$ is non-decreasing. By taking $k\rightarrow \infty, V^{t,k}_\beta(p)\rightarrow V^t_\beta(p)$.
Thus, $V^t_\beta(p)$ is a non-decreasing function.
\end{IEEEproof}

\begin{lemma}
\label{lm3}
Let $\Phi_{\hp}$ to be the set of values of $p_t$ for which offering an $\hp$ coupon is the optimal action at time $t$. Then, $\Phi_{\hp}$ is a convex set.
\end{lemma}
\begin{IEEEproof}
Since $\Phi_{\hp}=\{p\in[0,1],V^t_\beta(p)=V^t_{\beta, \hp}(p)\}$, assume that $p_t=ap_{t,1}+(1-a)p_{t,2}$ in which $p_{t,1},p_{t,2}\in \Phi_{\hp}$ and $a\in[0,1]$, $V^t_{\beta}(p_t)$ can be written as:
\begin{align} 
V^t_{\beta}(p_t)
& =V^t_{\beta}(ap_{t,1}+(1-a)p_{t,2})\\
&\ge aV^t_{\beta}(p_{t,1})+(1-a)V^t_{\beta}(p_{t,2})\\
&=aV^t_{\beta,\hp}(p_{t,1})+(1-a)V^t_{\beta,\hp}(p_{t,2})\\
&=a[(1-p_{t,1})[\beta^t\rhn+\beta V^t_{\beta}(\lambdana)]+p_{t,1}[\beta^t\rha+\beta V^t_{\beta}(\lambdaaa)]]\nonumber\\ &
\qquad +(1-a)[(1-p_{t,2})[\beta^t\rhn+\beta V^t_{\beta}(\lambdana)]+p_{t,2}[\beta^t\rha+\beta V^t_{\beta}(\lambdaaa)]]\\
&=V^t_{\beta,\hp}(ap_{t,1}+(1-a)p_{t,2}).
\end{align}
Thus, we have shown that: 
\begin{align}
V^t_{\beta}(p_t)\ge V^t_{\beta, \hp}(ap_{t,1}+(1-a)p_{t,1})=V^t_{\beta, \hp}(p_t).
\end{align}
By the definition of $V^t_{\beta}(p_t)$ in \eqref{eq:cphplp}, $V^t_{\beta}(p_t)\le V^t_{\beta, \hp}(p_t)$.
Therefore, $V^t_{\beta, \hp}(p_t)=V^t_{\beta}(p_t)$, which implies $\Phi_{\hp}$ is convex. 
\end{IEEEproof}
\subsection{Optimal Stationary Policy Structure}
\begin{theorem} 
\label{THSM}
There exists a threshold $\tau \in [0,1]$ such that the following policy is optimal:
\begin{align}
\pi^*(p_t)=
\begin{cases}
		\lp  & \mbox{if } \tau \le p_t \le 1 \\
		\hp & \mbox{if } 0 \le p_t\le \tau
\end{cases}
\label{eq:threshpolicy}
\end{align}
More precisely, 
assume $\delta\triangleq\rha-\rhn+\beta( V_\beta(\lambdaaa)- V_\beta(\lambdana))$, 
\begin{align}
\label{tau}
\tau=
\begin{cases}
\frac{\cl-(1-\beta)(\rhn+\beta V_\beta(\lambdana))}{(1-\beta)\delta} 
	& T(\tau)\geq\tau \\
\frac{\cl+\beta\lambdana(\rha+\beta V_\beta(\lambdaaa))}{{(1-(\lambdaaa-\lambdana)\beta)\delta}} -\frac{(1-\beta(1-\lambdana))(\rhn+\beta V_\beta(\lambdana))}{{(1-(\lambdaaa-\lambdana)\beta)\delta}}
	& T(\tau)<\tau
\end{cases}
\end{align}
where for $\lambdana\geq\tau$,
\begin{align}
V_\beta(\lambdana) &= V_\beta(\lambdaaa) = {\cl}/(1-\beta)
\end{align}
and for $\lambdana<\tau$,
\begin{align}
V_\beta(\lambdana) &= (1-\lambdana)[\rhn+V^1_{\beta}(\lambdana)] 
	\nonumber \\
	&\qquad \qquad +\lambdana[\rha+ V^1_{\beta}(\lambdaaa)], \\
V_\beta(\lambdaaa) &= \min\limits_{n\ge 0}\{G(n)\},
\end{align}
where\\
	\begin{align}
	G(n)&=\frac{\cl\frac{1-\beta^n}{1-\beta}+\beta^n[\bar{T}^n(\lambdaaa)(\rhn+C(\lambdana))+T^n(\lambdaaa)\rha]}{1-\beta^{n+1}[\bar{T}^n(\lambdaaa)\frac{\lambdana\beta}{1-(1-\lambdana)\beta}+T^n(\lambdaaa)]} \\
	T^n(\lambdaaa)&=\frac{(\lambdaaa-\lambdana)^{n+1}(1-\lambdaaa)+\lambdana}{1-(\lambdaaa-\lambdana)} \\
	\bar{T}^n(\lambdaaa) &=1-T^n(\lambdaaa) \\
	C(\lambdana) &=\beta\frac{(1-\lambdana)\rhn+\lambdana \rha}{1-(1-\lambdana)\beta}.
	\end{align}
\end{theorem} 

The proof of Theorem \ref{THSM} is provided in the Appendix~\ref{sec:app:threshthm} . An immediate consequence of this result is an upper bound on $p_t$ for offering an $\hp$ coupon.

We define $\kappa$ to be the ratio between the gain from offering an $\hp$ coupon to a $\normal$ consumer and the loss from offering an $\hp$ coupon to a consumer whom the retailer thinks is $\normal$ but is actually $\alerted$. Thus, 	
\begin{align}
	\kappa = \frac{\cl-\rhn}{\rha-\rhn}.
\end{align} For fixed costs, the threshold can be bounded by the following two Corollaries.

\begin{corollary}
\label{COR1}In the model where transition probabilities $(\lambdana, \lambdaaa)$ are unknown to the retailer, if 
	\begin{align}
	\label{eq:unk:phpopt}
	p_t\le \kappa,
	\end{align}
then it is optimal for the retailer to offer an $\hp$ coupon.
\end{corollary}
\begin{corollary}
\label{CORBD}
 Fix the costs and $\lambdaaa$, let $\lambda_1=\frac{\cl-\rhn}{\rha-\rhn}$ and $\lambda_2$ be the solution of $\frac{\lambda_2}{1-(\lambdaaa-\lambda_2)}=\frac{\beta(\cl-\rha)\lambda_2+\cl-\rhn}{(1-\beta)\rha-\rhn+\beta\cl}$. When $\lambdana\ge\lambda_2$, the threshold $\tau$ in the optimal stationary policy can be written as a closed form expression with respect to $\lambdana$:
 if $\lambdana>\lambda_1$, 
 \begin{align}
 \tau=\kappa;
 \end{align}
 if $\lambda_2<\lambdana<\lambda_1$, 
 \begin{align}
 \tau=\frac{\beta(\cl-\rha)\lambdana+\cl-\rhn}{(1-\beta)\rha-\rhn+\beta\cl}.
 \end{align}
 Moreover, if $\lambdana<\lambda_2$, $\tau$ can be upperbounded by 
\begin{align}
\bar{\tau} = \frac{\lambda_2}{1-(\lambdaaa-\lambda_2)}.
\end{align}
\end{corollary}
\begin{figure}[h]
\centering
\includegraphics[width=80mm]{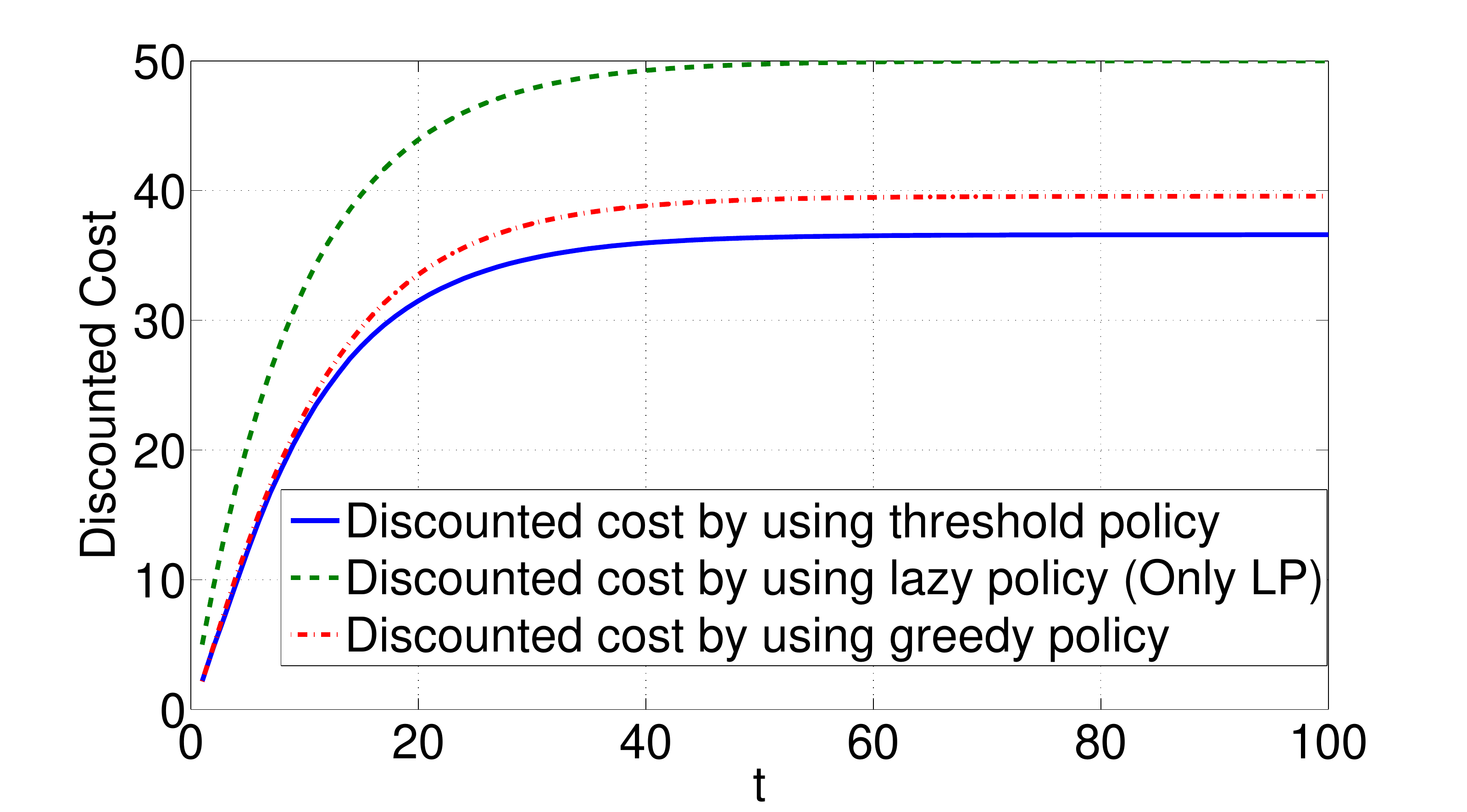}
\caption{Discounted cost resulted by using different decision policies}
\label{fig:RWE}
\end{figure}
A detailed proof of Corollary~\ref{COR1} and~\ref{CORBD} are presented in the Appendix~\ref{PFCOR1} and Appendix~\ref{PFCORBD}, respectively. 
\begin{figure}[h]
 \centering
    \subfloat[Threshold $\tau$ vs. $\lambdana$. (Parameters:  $\beta=0.9, \cl=3,\rhn=1,\rha=12, \kappa=0.18.$)]{{\includegraphics[width=58mm]{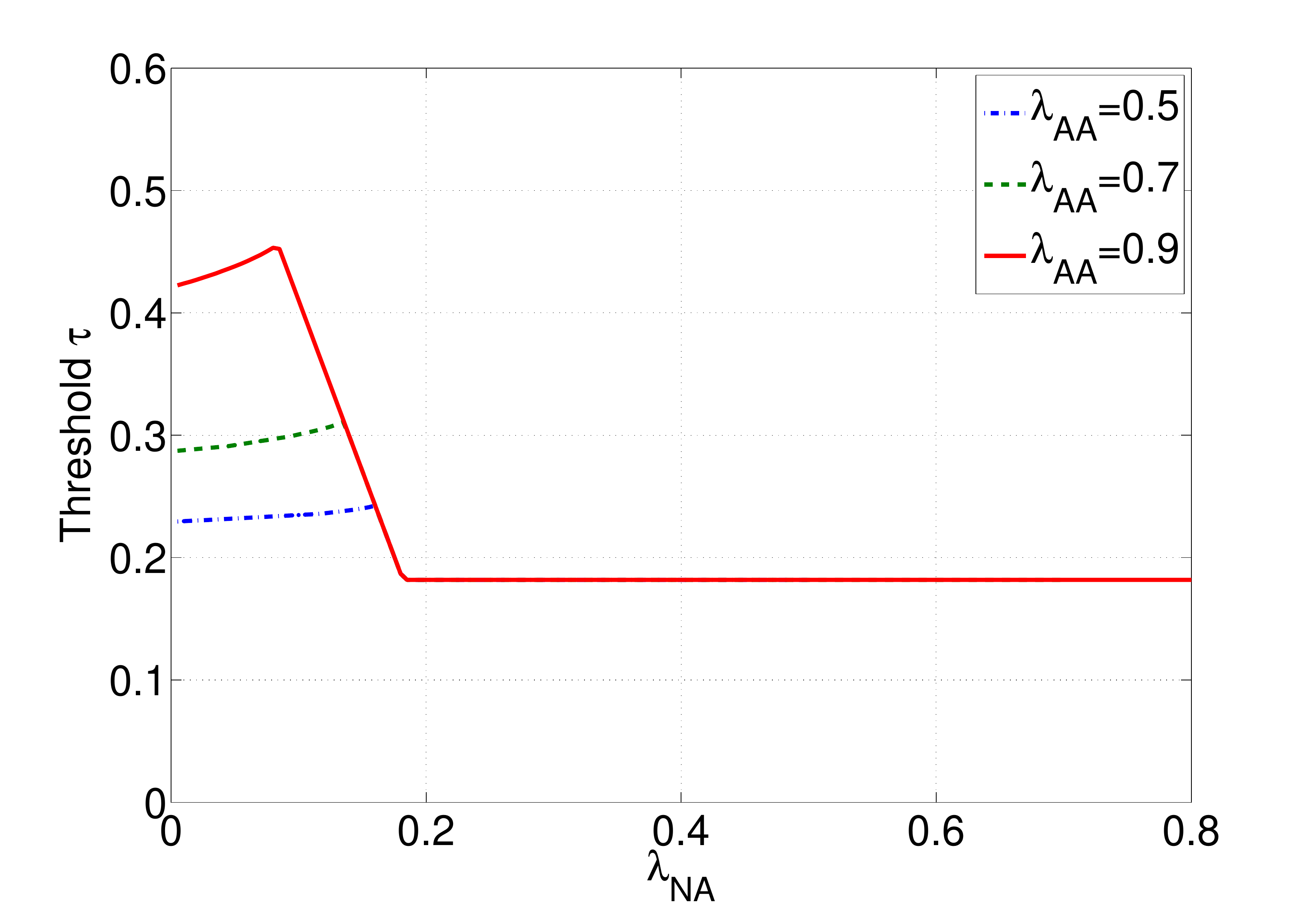} }\label{THP1}}%
    \qquad
    \subfloat[Threshold $\tau$ vs. $\lambdana$. (Parameters: $\lambdaaa=0.7, \beta=0.9, \rhn=1,\rha=12.$)]{{\includegraphics[width=58mm]{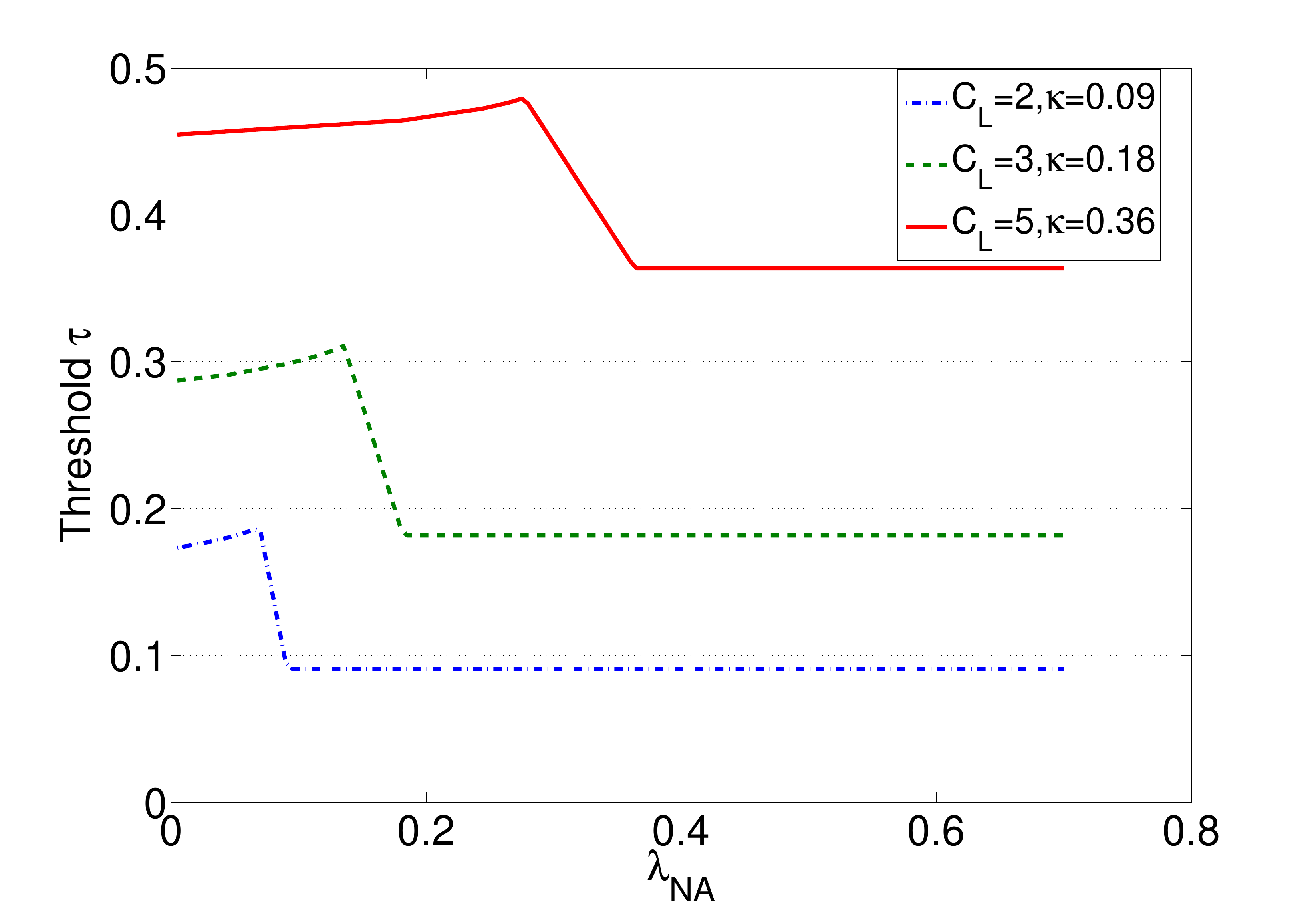} }\label{THP2}}%
    \caption{Threshold $\tau$ vs. $\beta$ for different values of $\lambdaaa$ and $\lambdana$ }%
    \label{fig:figure5}%
\end{figure}
To illustrate the performance of the proposed threshold policy, we compare the discounted cost resulted from the threshold policy with the greedy policy which minimize the instantaneous cost at each decision epoch as well as a lazy policy which a retailer only offers $\lp$ coupons. We plot the discounted cost averaged over 1000 independent MDPs w.r.t. time $t$ for different decision policies in Fig.~\ref{fig:RWE}. The illustration demonstrates that the proposed threshold policy performs better than the greedy policy and the lazy policy. 

Figure~\ref{THP1} shows the optimal threshold policy with respect to $\lambdana$ for three fixed choices of $\lambdaaa$. It can be seen that the threshold is increasing when $\lambdana$ is small, this is because for a small $\lambdana$, the \consumer s is less likely to transition from $\normal$ to $\alerted$. Therefore, the retailer tends to offer an $\hp$ coupon to the \consumer. When $\lambdana$ gets larger, the \consumer\ is more likely to transition from $\normal$ to $\alerted$. Thus, the retailer tends to play conservatively by decreasing the threshold for offering an $\lp$ coupon. When $\lambdana$ is greater than $\kappa$, the retailer will just use $\kappa$ to be the threshold for offering an $\hp$ coupon. One can also observe that with increasing $\lambdaaa$, the threshold $\tau$ decreases. On the other hand, for fixed $\rhn$ and $\rha$, Figure~\ref{THP2} shows that the threshold $\tau$ increases as the cost of offering an $\lp$ coupon increases, making it more desirable to take a risk and offer an $\hp$ coupon. 

\begin{figure}[h]%
    \centering
    \subfloat[Threshold $\tau$ vs. $\beta$ for different values of $\lambdaaa$ (Parameters: $\lambdana=0.1, \cl=3, \rhn=1,\rha=12, \kappa=0.18.$)]{{\includegraphics[width=58mm]{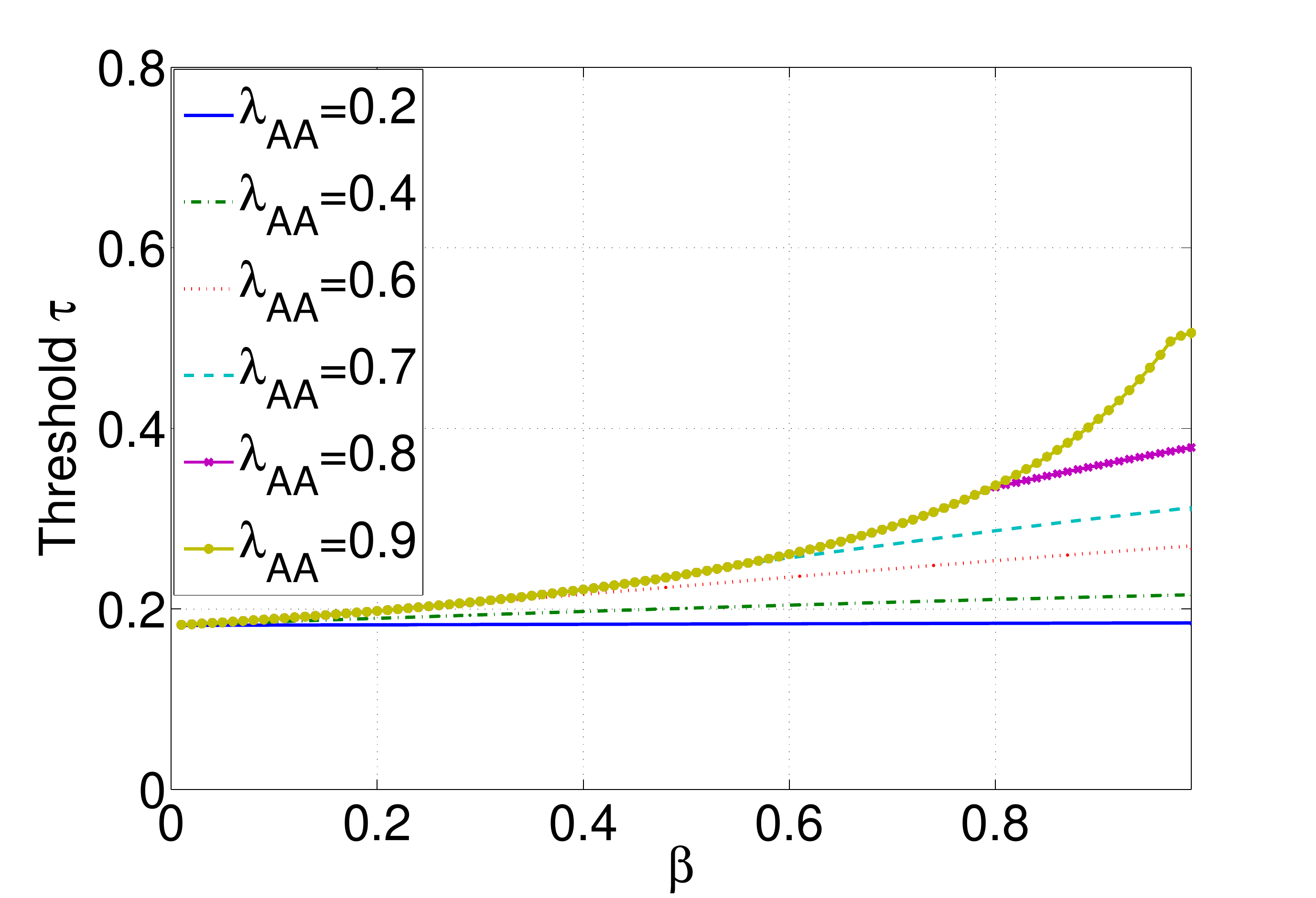} }\label{Tvsbeta1}}%
    \qquad
    \subfloat[Threshold $\tau$ vs. $\beta$ for different values of $\lambdana$ (Parameters: $\lambdaaa=0.7, \cl=3, \rhn=1,\rha=12.$)]{{\includegraphics[width=58mm]{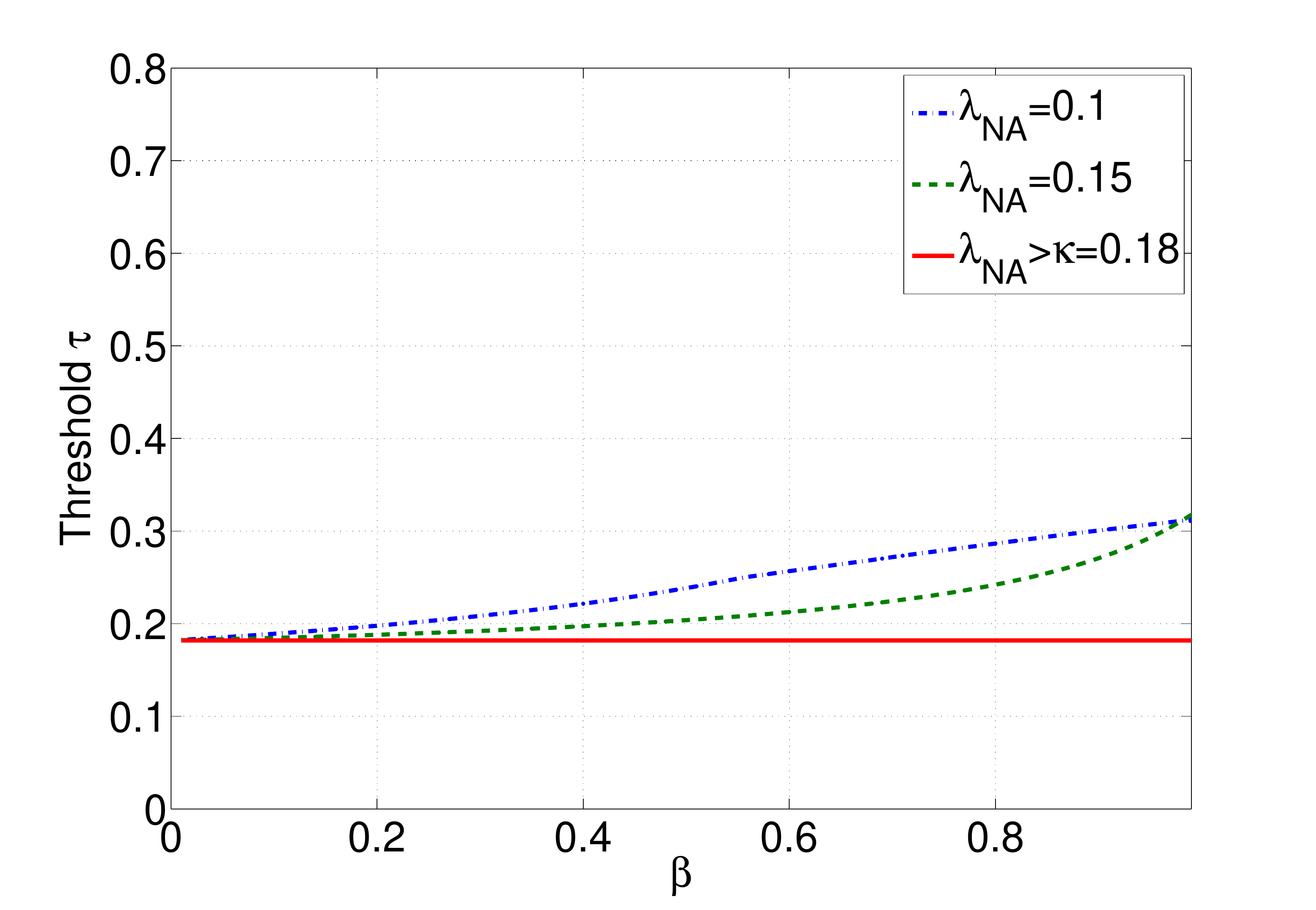} }\label{Tvsbeta2}}%
    \caption{Threshold $\tau$ vs. $\beta$ for different values of $\lambdaaa$ and $\lambdana$ }%
    \label{fig:figure11}%
\end{figure}
\begin{figure}[h]
\centering
\includegraphics[width=80mm]{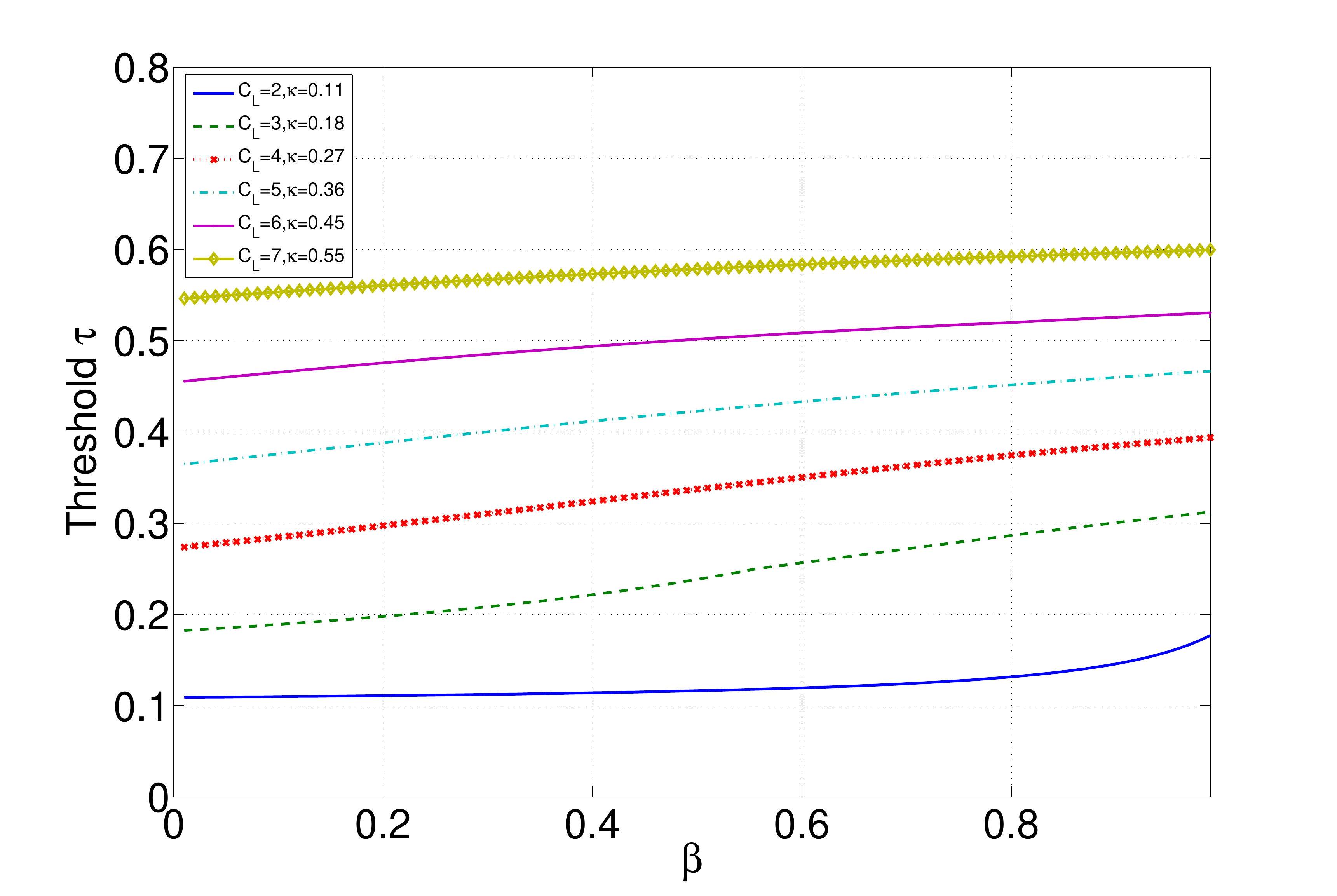}
\caption{Threshold $\tau$ vs. $\beta$ for different values of $\cl$. (Parameters: $\lambdana=0.1, \lambdaaa=0.9, \rhn=1,\rha=12.$)}
\label{fig:figure12}
\end{figure}
The relationship between the discount factor $\beta$ and the threshold $\tau$ as functions of transition probabilities is shown in Figure~\ref{fig:figure11}. It can be seen in Figure~\ref{Tvsbeta1} that the threshold increases as $\beta$ increases. This is because when $\beta$ is small, the retailer values the present rewards more than future rewards. Therefore, the retailer tends to play conservatively so that it will not ``creep out" the \consumer\ in the present. Figure~\ref{Tvsbeta2} shows that the threshold is high when $\lambdaaa$ is large or $\lambdana$ is small. A high $\lambdaaa$ value indicates that a \consumer\ is more likely to remain in $\alerted$ state. The retailer is willing to play aggressively since once the \consumer\ is in alerted state, it can take a very long time to transition back to $\normal$ state. A low $\lambdana$ value implies that the \consumer\ is not very privacy sensitive. Thus, the retailer tends to offer $\hp$ coupons to reduce cost. One can also observe in Figure~\ref{Tvsbeta2} that the threshold $\tau$ equals to $\kappa$ after $\lambdana$ exceeds the ratio $\kappa$. This is consistent with results shown in Figure~\ref{fig:figure5}.

The effect of an $\lp$ coupon cost on the threshold for different discount factors is plotted in Figure~\ref{fig:figure12}. It can be seen that a higher $\cl$ will increase the threshold because the retailer is more likely to offer an $\hp$ coupon when the cost of offering an $\lp$ coupon is high.

\subsection{Consumer with Multi-Level Alerted States}

In this section, we study the case that the \consumer\ has multiple $\alerted$ states. 
Without loss of generality, we define the transition matrix to be
\begin{equation}
\label{TMtrx3}
\mathbf{\Lambda} =
 \begin{pmatrix}
  \lambda_{N,N} & \lambda_{N,A_1} & \dots & \lambda_{N,A_K} \\
  \lambda_{A_1,N} & \lambda_{A_1,A_1} & \dots & \lambda_{A_1,A_K} \\
  \vdots & \vdots & \ddots & \vdots \\
  \lambda_{A_K,N} & \lambda_{A_K,A_1} & \dots & \lambda_{A_K,A_K}
 \end{pmatrix}
\end{equation} and $\bar{\mbf{e}}_i$ to be the $i^{th}$ row of $\mbf{\Lambda}$. The expected cost at time $t$, given belief $\mbf{\bar{p}}_t$ and action $u_t$, has the following expression:
\begin{align}
C(\bar{\mbf{p}}_t,u_t)=
\left\{
	\begin{array}{ll}
		\cl  & \mbox{if } u_t=\lp \\
		\mbf{\bar{p}}_t^{T}\bar{\mbf{C}} & \mbox{if } u_t=\hp
	\end{array}
\right..
\end{align}

Assuming that the retailer has perfect information about the belief states, the cost function evolves as follows. By using an $\lp$ coupon at time $t$,
\begin{align}
V^t_{\beta, \lp}(\bar{\mbf{p}}_t)=\beta^t\cl+V^{t+1}_\beta(\bar{\mbf{p}}_{t+1})=\beta^t\cl+V^{t+1}_\beta(T(\bar{\mbf{p}}_t)),
\end{align}
where $T(\bar{\mbf{p}}_{t})=\bar{\mbf{p}}_{t}^T\mbf{\Lambda}$ is the Markov transition operator generalizing \eqref{eq:T}. By using an $\hp$ coupon at time $t$,
\begin{align}
V^t_{\beta, \hp}(\bar{\mbf{p}}_t)=\beta^t\bar{\mbf{p}}_t^{T}\bar{\mbf{C}}+V^{t+1}_\beta(\bar{\mbf{p}}_{t+1})=\beta^t\bar{\mathbf{p}}_t^{T}\bar{\mathbf{C}}+\bar{\mathbf{p}}_t^{T} 
\left( \begin{array}{c}
V_{\beta}^{t+1}(\bar{\mathbf{e}}_1)  \\
V_{\beta}^{t+1}(\bar{\mathbf{e}}_2)  \\
\vdots\\
V_{\beta}^{t+1}(\bar{\mathbf{e}}_{K+1})  \end{array} \right).
\end{align}
Therefore, by~\eqref{eq:cphplp}, we have
$V^t_\beta(\bar{\mbf{p}}_t)=\min\{V^t_{\beta, \lp}(\bar{\mbf{p}}_t),V^t_{\beta, \hp}(\bar{\mbf{p}}_t)\}.$

\begin{figure}
  \centering
  \includegraphics[width=76mm]{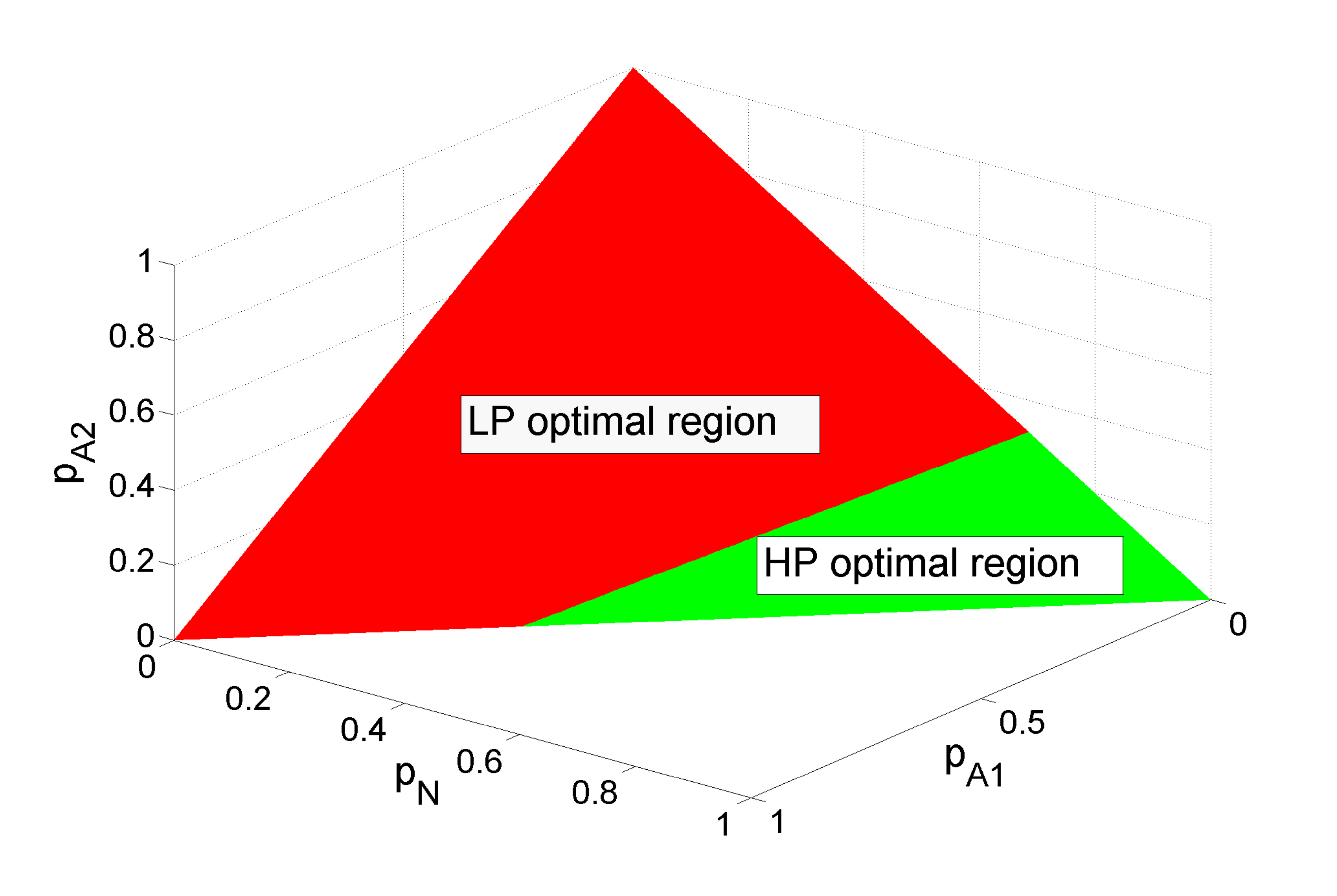}
  \caption{Example of the optimal policy region for three-state consumer. (Parameters: $\lambda_{\msf{N},\msf{N}}=0.7, \lambda_{\msf{N},A1}=0.2, \lambda_{\msf{N},\msf{A2}}=0.1;
    \lambda_{\msf{A1},\msf{N}}=0.2, \lambda_{\msf{A1},\msf{A1}}=0.5, \lambda_{\msf{A1},\msf{A2}}=0.3;
    \lambda_{\msf{A2},\msf{N}}=0.1, \lambda_{\msf{A2},\msf{A1}}=0.2, \lambda_{\msf{A2},\msf{A2}}=0.7;\beta=0.9,C_{\msf{L}}=7,C_{\msf{HN}}=1,C_{\msf{HA1}}=10,C_{\msf{HA2}}=20$).}
  \label{fig:figure3s}
\end{figure}
In this problem, since the instantaneous costs are nondecreasing with the state when the action is fixed and the evolution of belief state is the same for both $\lp$ and $\hp$, the existence of an optimal stationary policy with threshold property is guaranteed by Proposition 2 in~\cite{lovejoy1987some}. 
The optimal stationary policy for a three-state \consumer\ model is illustrated in Figure \ref{fig:figure3s}. For fixed costs, the plot shows the partition of the belief space based on the optimal actions and reveals that offering an $\hp$ coupon is optimal when $p_{N,t}$, the belief of the \consumer\ being in $\normal$ state, is high.
 


\section{Consumers with Coupon Dependent Transitions}
\label{sec:cpdep}
Generally, \consumer s' reaction to $\hp$ and $\lp$ coupons are different. To be more specific, a \consumer\ is likely to feel less comfortable when being offered a coupon on medication ($\hp$) than food ($\lp$). Thus, we assume that the Markov transition probabilities are dependent on the coupon offered. Let $p_t$ denote the belief of a \consumer\ being in the $\alerted$ state at time $t$. 

As shown in Figure~\ref{fig:figure3}, by offering an $\lp$ coupon, the state transition follows the Markov chain
\begin{align}
\mbf{\Lambda}_{\lp} =
  \begin{pmatrix}
   1-\lambdana & \lambdana \\
   1-\lambdaaa & \lambdaaa 
  \end{pmatrix}.
\end{align} 
Otherwise, the state transition follows
\begin{align}
\mbf{\Lambda}_{\hp} =
  \begin{pmatrix}
   1-\lambdapna & \lambdapna \\
   1-\lambdapaa & \lambdapaa 
  \end{pmatrix}.
\end{align}
According to the model in Section~\ref{sec:sysmod},  $\lambdaaa>\lambdana,\lambdaaa'>\lambdana'$. Moreover, we assume that offering an $\hp$ coupon will increase the probability of transition to or staying at $\alerted$ state. Therefore, $\lambdaaa'>\lambdaaa$ and $\lambdana'>\lambdana$. The minimum cost function evolves as follows: for an $\hp$ coupon offered at time $t$, we have
\begin{align*}
V^t_{\beta, \hp}(p_t)=\beta^tC(p_t,\hp)+(1-p_t)V^{t+1}_\beta(\lambdana') +p_tV^{t+1}_\beta(\lambdaaa').
\end{align*} Otherwise,
\begin{align*}
V^t_{\beta, \lp}(p_t)
&=\beta^t\cl+V^{t+1}_\beta(p_{t+1})=\beta^t\cl+V^{t+1}_\beta(T(p_t)),
\end{align*}
where $T(p_t)=\lambdana(1-p_t)+\lambdaaa p_t$ is the one step transition defined in Section~\ref{sec:sysmod}. 
\begin{theorem}
\label{THMCDT}
Given action dependent transition matrices $\mathbf{\Lambda_{\lp}}$ and $\mathbf{\Lambda_{\hp}}$, the optimal stationary policy has threshold structure. 
\end{theorem} 
\begin{figure}
  \centering
  \includegraphics[width=130mm]{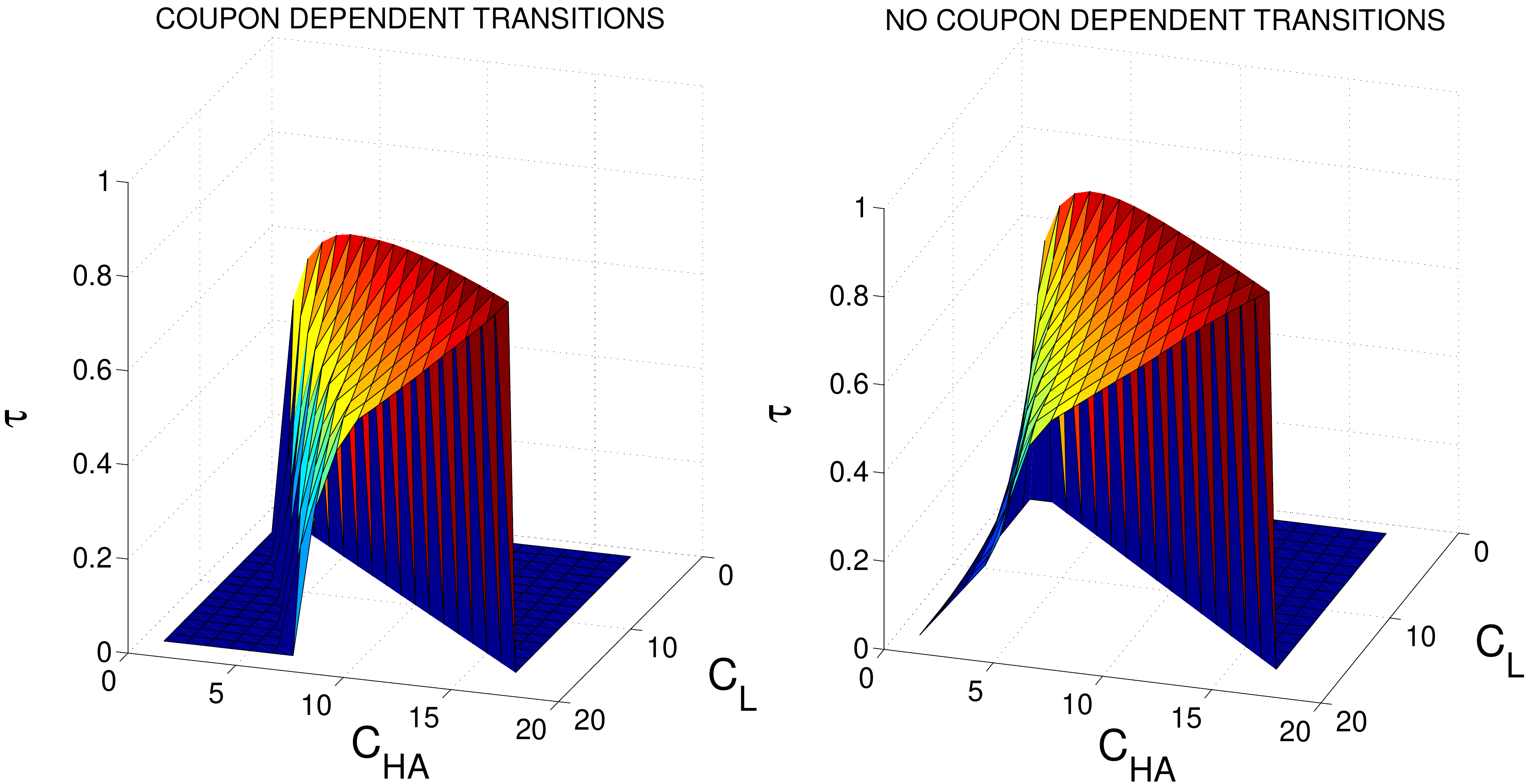}
  \caption{Optimal policy threshold for consumer with/without coupon dependent transition probabilities. (Parameters: $\lambdana=0.2,\lambdaaa=0.8,\lambdana'=0.5,\lambdaaa'=0.9,\beta=0.9$).}
  \label{fig:figure6}
\end{figure}
A detailed proof of Theorem~\ref{THMCDT} is presented in the Appendix \ref{sec:app:cpdthm}.

Figure~\ref{fig:figure6} shows the effect of costs on the threshold $\tau$. We can see that for a fixed ${\cl}$ and  ${\rha}$ pair, the threshold for $\lp$ coupons for \consumer s in this model is lower than our original model without coupon-dependent transition probabilities. The retailer can only offer an $\lp$ coupon with certain combination of costs; we call this the $\lp$-only region. One can also see that the $\lp$-only region for the coupon-independent transition case is smaller than that for the coupon-dependent transition case since for the latter, the likelihood of being in an $\alerted$ state is higher for the same costs. 

\section{Policies under Noisy Cost Feedback and Uncertain Initial Belief}
\label{sec:bayes}

In this section, we study the case in which the received costs are random. In the previous sections, if the retailer offered an $\hp$ coupon at time $t$, then it could learn the state of the consumer at time $t$ based on whether there received cost was $C_{HN}$ or $C_{HA}$. If the cost feedback is random, the the retailer may not be able to infer the consumer's state exactly. We describe  policy heuristics for this setting that perform Bayesian estimation of the quantity $p_t$ used in the threshold policy earlier. This approach is also useful when the initial value $p_0$ is not known to the retailer.

We model the noisy cost feedback by assuming the received cost $C_t$ is random. The distribution of $C_t$ is given by a conditional probability density $f( c | G_t, u_t)$ on a bounded subset of $\mathbb{R}$, where $G_t$ is the state of the consumer and $u_t$ is the action taken by the retailer at time $t$.  To match the previous model, we further take $f(c | G_t = \alerted, u_t = \lp) = f( c | G_t = \normal, u_t = \lp)$ to indicate that the received cost conveys no information about the state under an $\lp$ coupon. Let $f(c | u_t = \lp) = f(c | G_t = \alerted, u_t = \lp)$. For a given value $p_t = p$, define the likelihood of observing a cost $C_t = c$ under the two coupons:
	\begin{align}
	\ell( c | \lp, p) &= f(c | \alerted, \lp) 
		\label{eq:bayes:lplikelihood} \\
	\ell( c | \hp, p) &= f(c | \normal, \hp) (1-p) + f(c | \alerted, \hp) p
		\label{eq:bayes:hplikelihood}
	\end{align}
These likelihoods will be useful in defining the two estimators.

In both approaches in this section the retailer computes an estimate $\hat{p}_t$ of the probability $p_t$ that $G_t = \alerted$. It then uses \eqref{eq:threshpolicy} to decide which coupon to offer at time $t$ by comparing $\hat{p}_t$ to a version of the threshold in \eqref{tau}. Define $\mathcal{C_L},\mathcal{C_{HN}}$, and $\mathcal{C_{HA}}$ to be the feasible cost sets $\{ c : f( c | \lp ) > 0 \}$, $\{c : f( c | \alerted, \hp) > 0\}$, and $\{ c : f( c | \normal, \hp) > 0 \}$, respectively. Since $\tau$ involves the costs $\cl$, $\rhn$ and $\rha$, there are several ways to compute an approximate threshold under the cost uncertainty. 

Firstly, we can set $\cl$, $\rhn$ and $\rha$ to be the expected costs:
	\begin{align}
	\cl &= \int_{\mathbb{R}} c f( c | \lp ) dc
	\\
	\rhn & = \int_{\mathbb{R}} c f( c | \normal, \hp) dc
	\\
	\rha &= \int_{\mathbb{R}} c f( c | \alerted, \hp) dc.
	\end{align}
Plugging these into \eqref{tau} gives the mean threshold $\tau_{\mathsf{avg}}$.  Since $\tau$ is monotonically increasing in $\cl$ and $\rha$ and monotonically decreasing in $\rhn$, we can compute and upper bound on $\tau$ by setting $\cl = \max\{ c : c\in\mathcal{C_L} \}$, $\rha = \max\{c : c\in\mathcal{C_{HA}}\}$, and $\rhn = \max\{ c : c\in\mathcal{C_{HN}} \}$. These values give the upper bound threshold $\tau_{\mathsf{max}}$. Similarly, by setting $\cl$ and $\rha$ to the lower bounds on the support and $\rhn$ to the upper bound, we obtain a lower bound threshold $\tau_{\mathsf{min}}$. Finally, we computed a robust version of threshold $\tau_{\mathsf{R}}$ as $ \tau_{\mathsf{R}} = \{\tau: \max \limits_{\cl,\rhn,\rha} \{\min \limits_{\pi(p_t)} V^t_{\beta}(p_t)\}\}$, where $(\cl,\rhn,\rha)\in \mathcal{C_L}\times\mathcal{C_{HN}}\times\mathcal{C_{HA}}$, is the  This threshold policy is the largest (cost case) threshold over all possible combination of costs. Thus, it gives the $max-min$ value of the total discounted cost. We can see that the total discounted cost induced by this robust version of threshold is close to that induced by using the upper bound of costs. 

\subsection{MAP Estimation of the Consumer State}
\label{sec:mapcs}
In the previous model, if $u_t = \hp$ the retailer could infer $G_t$ based on $C_t$, so $p_{t+1}$ is given by the state transitions of the Markov chain. With noisy costs this exact inference is no longer possible. A simple heuristic for the retailer is to try to infer $G_t$ based on the random cost $C_t$, compute an estimate of $p_t$, and then use the previous strategy.

At time $t = 1$, given an initial $p_0$ we estimate $\hat{p}_1 = T(p_0)$. The retailer then applies the threshold policy \eqref{eq:threshpolicy} with input $\hat{p}_1$ to offer a coupon. For times $t = 2, 3, \ldots$ the retailer treats the estimate $\hat{p}_{t-1}$ as an estimate of the probability that $G_{t-1} = \alerted$. 
If $u_{t-1} = \lp$, then the retailer sets $\hat{p}_t = T(\hat{p}_{t-1})$. If $u_{t-1} = \hp$ then the retailer uses a maximum a posteriori probability (MAP) detection rule to estimate the state $G_{t-1}$ based on the received cost $C_{t-1}$. That is, it sets $\hat{G}_{t-1} = \normal$ if
	\begin{align}
	\frac{f( C_{t-1} | \normal, \hp) (1 - \hat{p}_{t-1})
		}{
		f( C_{t-1} | \alerted, \hp) \hat{p}_{t-1} 
		} > 1 
	\end{align}
and $\hat{G}_{t-1} = \alerted$ otherwise, where $C_{t-1}$ is the received cost at time $t-1$. It then uses the following estimate $p_t$ at time $t$:
	\begin{align}
	\hat{p}_{t} = 
		\begin{cases}
		\lambdana  & \mbox{if } \hat{G}_t=\normal \\
		\lambdaaa  & \mbox{if } \hat{G}_t=\alerted
		\end{cases}
	\end{align}

\begin{figure}
  \centering
  \includegraphics[width=90mm]{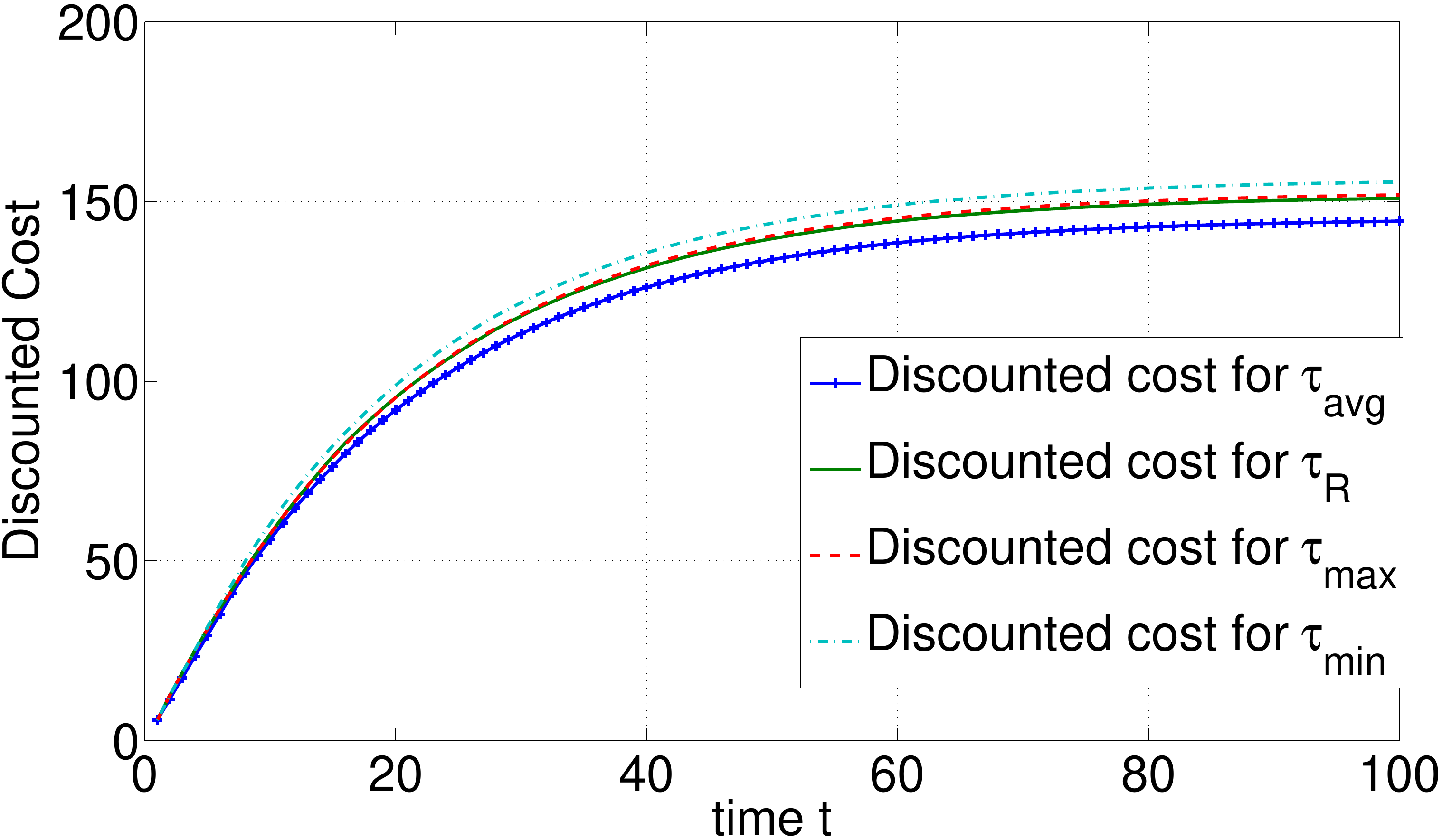}
  \caption{Temporal discounted costs for different heuristics on computing thresholds. (Parameters: $\lambdana=0.2$, $\lambdaaa=0.8$, $p_0=0.2$,$\beta=0.95$, $f(c | \lp) = \unif[6,10]$, $f(c | \normal, \hp) = \unif [0.2,5.8]$, and $f(c | \alerted, \hp) = \unif[12,20]$). The discounted cost is averaged over 1000 independent runs.}
  \label{fig:mrbss}
\end{figure}

Essentially, the retailer uses MAP estimation to infer $G_{t-1}$ after receiving the cost $C_{t-1}$ from the action $u_{t-1} = \hp$. If the densities $f( c | \normal, \hp)$ and $f(c | \alerted, \hp)$ have disjoint supports, then the inference of $G_{t-1}$ is error free, so $\hat{G}_{t-1} = G_{t-1}$ and the estimate $\hat{p}_t$ is correct. Figure~\ref{fig:mrbss} shows the discounted cost as a function of time for some different variants of the threshold in \eqref{tau}. In this example the cost distributions are uniformly distributed in disjoint intervals.  The plot shows that the mean threshold yields a total discounted cost that is slightly less than the upper and lower bound thresholds.

\subsection{Bayesian Estimation of State Probabilities}

In the previous approach, the retailer estimates the underlying state and then uses this to form an estimate of the probability $p_t$ that $G_t = \alerted$. A different approach is to form a Bayes estimate of $p_t$: the retailer computes a probability distribution on $[0,1]$ representing its uncertainty about $p_t$. To choose an action $u_t$ it can use a point estimate of $p_t$ to use in \eqref{eq:threshpolicy} with one of the thresholds described before.

In this formulation, the estimator of $p_t$ is a probability distribution. Let $q_{t-1}( p )$ be the estimator of $p_{t-1}$. The retailer treats this as a prior distribution. Upon receiving the cost $C_{t-1}$ it computes a posterior estimate on $p_{t-1}$ using Bayes rule. If $u_{t-1} = \hp$, it sets
	\begin{align}
	q_{t-1}( p | C_{t-1}) = \frac{ 
		\ell( C_{t-1} | \hp, p) q_{t-1}(p)
		}{
		\int_{0}^{1} \ell( C_{t-1} | \hp, p') q_{t-1}(p')	 dp'
		}
	\end{align}
If $u_{t-1} = \lp$ then from \eqref{eq:bayes:lplikelihood} we can see that $\ell( C_{t-1} | \lp, p)$ does not depend on $p$, so the posterior $q_{t-1}( p | C_{t-1}) = q_{t-1}( p )$ in this case. Given the posterior estimate $q_{t-1}( p | C_{t-1})$ the retailer then evolves the state distribution through the Markov chain governing the state to form the prior distribution $q_t(p)$ for estimating $p_t$ at time $t$. That is, if $P_{t-1}$ is a random variable with distribution $q_{t-1}( p | C_{t-1})$, then $q_t(p)$ is the distribution of $T(P_{t-1})$. Let $Q_{t-1}(p | C_{t-1}) = \int_{0}^{p} q_{t-1}( p' | C_{t-1})$ be the cumulative distribution function of $P_{t-1}$. Then
	\begin{align}
	\mathbb{P}\left( T(P_{t-1}) \le p \right) 
	= \mathbb{P}\left( P_{t-1}  \le \frac{ p - \lambdana}{ \lambdaaa - \lambdana} \right) 
	= Q_{t-1}\left( \frac{ p - \lambdana}{ \lambdaaa - \lambdana} \big| C_{t-1} \right)
	\end{align}
so
	\begin{align}
	q_t(p) = \frac{1}{ \lambdaaa - \lambdana } q_{t-1}\left( \frac{ p - \lambdana}{ \lambdaaa - \lambdana } \big| C_{t-1} \right).
	\end{align}

The retailer then uses $q_{t}(p)$ to form a point estimate $\hat{p}_t$ of $p_t$ suitable for applying the threshold policy in \eqref{eq:threshpolicy} and \eqref{tau}. We consider two such point estimates which we call the mean and max estimators, respectively:
	\begin{align}
	\hat{p}_{t,\mathsf{mean}} &= \int_{0}^{1} p q_t(p) dp \\
	\hat{p}_{t,\mathsf{MAP}} &= \argmax_{p \in [0,1]} q_t(p).
	\end{align}
\begin{figure}
  \centering
  \includegraphics[width=90mm]{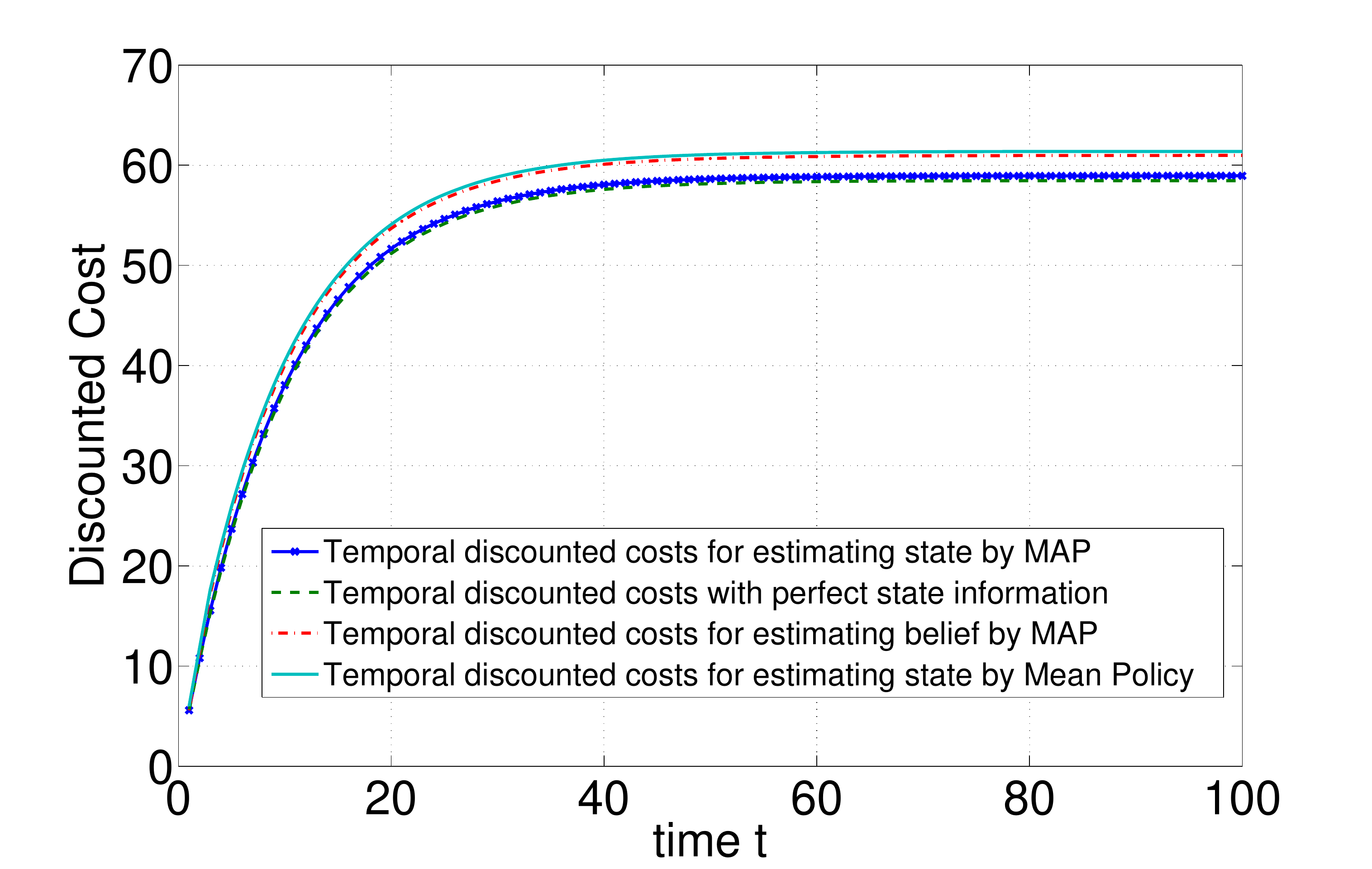}
  \caption{Temporal discounted costs for different estimation mechanisms. (Parameters: $\lambdana=0.2$,$\lambdaaa=0.8$, $p_0=0.2$,$\beta=0.9$, $f(c | \lp) = \unif[3,9]$, $f(c | \normal, \hp) = \unif[0.25,7.75]$, $f(c | \alerted, \hp) = \unif[6,18]$). The discounted cost is averaged over 1000 independent runs.}
  \label{fig:estjc}
\end{figure}
Figure \ref{fig:estjc} shows the discounted cost versus time for uniformly distributed costs with overlapping support. The decision is made by following the optimal stationary policy computed by the mean threshold in \ref{sec:bayes}. We illustrate the result for four algorithms: the solid curve and the dash-dot curve are the MAP and mean strategy described above, respectively; the dashed curve is a policy in which the costs are random but the algorithm is given side information about $G_{t}$ after choosing $u_t = \hp$ (perfect state information); finally, the curve with cross is the MAP estimate of actual state $G_t$ described in Section~\ref{sec:mapcs}. In this example, as one can expect, decision making with perfect state information has the minimum discounted cost. MAP estimation of $G_t$ results in an 0.82$\%$ increase in total discounted cost compared to the case in which the retailer receives perfect information about \consumer\ state. However, the MAP and mean policy to estimate belief state $p_t$ only have 2.9$\%$ and 4.29$\%$ increase, respectively. Thus, the MAP for estimating belief perfoms slightly better than the Mean policy. Effectively, the lack of initial belief knowledge does not affect the discouted cost very much on average. This is because offering an $\hp$ coupon allows the retailer to learn the actual state from the cost feedback, thus, reset the belief state. 
\section{Conclusions}
\label{sec:conclusion}
We proposed a POMDP model to capture the interactions between a retailer and a privacy-sensitive consumer in the context of personalized shopping. The  retailer seeks to minimize the expected discounted cost of violating the \consumer's privacy. We showed that the optimal coupon-offering policy is a stationary policy that takes the form of an explicit threshold that depends on the model parameters. In summary, the retailer offers an $\hp$ coupon when the $\normal$ to $\alerted$ transition probability is low or the probability of staying in $\alerted$ state is high. Furthermore, the threshold optimal policy also holds for \consumer s whose privacy sensitivity can be captured via multiple alerted states as well as for the case in which \consumer s exhibit coupon-dependent transition. For the case in which the cost feedbacks from the consumer are noisy, we have introduced a heuristic method using the mean value of costs to compute the decision threshold. Furhtermore, under noisy cost feedbacks scenario, we have introduced a Bayesian data analysis approach for decision making which includes estimating \consumer\ belief state when the initial belief state is unknown to the retailer. Our work suggests several interesting future directions: one straightfoward extension of our work is to model uncertainties in the statistical model for the \consumer\ transition probabilities. Further a field, one can also develop game theoretic models to study the interaction between a retailer and strategic consumers and develop methods to test those models in practice.

\appendices

\section{Proof of Theorem \ref{THSM} \label{sec:app:threshthm}}

\begin{IEEEproof}
Let $p_F$ be the stationary distribution of the Markov transition. Then $p_F=\lambdaaa p_F+(1-p_F)\lambdana$, which implies $p_F=\frac{\lambdana}{1-\lambdaaa+\lambdana}$.
Remember that the threshold is the solution to $V^t_{\beta,\lp}(p_t)=V^t_{\beta,\hp}(p_t)$. Let $\tau$ be the threshold value, we have:
\begin{align}
\label{TSequal}
\begin{split}
&\beta^t\cl+ V^{t+1}_\beta(T(\tau))\\
&=(1-\tau)[\beta^t\rhn+ V^{t+1}_{\beta}(\lambdana)]+\tau[\beta^t\rha+ V^{t+1}_{\beta}(\lambdaaa)].
\end{split}
\end{align}

By the definition of $V^{t}_\beta(p_t)$, we know that $V^{t}_\beta(p_t)=\beta^t V_\beta(p_t)$. Thus $V^{t}_\beta(\lambdana)=\beta^t V_\beta(\lambdana)$ and $V^{t}_\beta(\lambdaaa)=\beta^t V_\beta(\lambdaaa)$. 

If $T(\tau)\ge \tau$, which is equivalent to $p_F\ge \tau$, then $V^{t+1}_{\beta}(T(\tau))=V^{t+1}_{\beta,\lp}(T(\tau))$. Therefore, $V^t_{\beta,\lp}(\tau)=\lim\limits_{n\rightarrow\infty}\{\beta^t\frac{1-\beta^{n}}{1-\beta}\cl+\beta^nV^{t+1}_\beta(T^n(\tau))\}$ where $T^n(\tau)=T(T^{n-1}(\tau))=p_F(1-(\lambdaaa-\lambdana)^n)+(\lambdaaa-\lambdana)^n\tau$. Taking $n\rightarrow \infty$, we have $V^t_{\beta,\lp}(\tau)=\beta^t\frac{C}{1-\beta}$. Substitute this into \eqref{TSequal} yields: 
\begin{align}
\frac{\cl}{1-\beta}=(1-\tau)\rhn+\tau \rha+\beta(\tau V_\beta(\lambdaaa)+(1-\tau)V_\beta(\lambdana)).
\end{align}
By rearranging terms in the above expression, we have
\begin{align}
\label{eq:tau1}
\tau=\frac{\frac{\cl}{1-\beta}-\rhn-\beta V_\beta(\lambdana)}{(\rha-\rhn)+\beta(V_\beta(\lambdaaa)-V_\beta(\lambdana))}.
\end{align}

If $p_F \le \tau $, then $T(\tau)\le \tau$. Therefore  $V^{t+1}_{\beta}(T(\tau))=V^{t+1}_{\beta,\hp}(T(\tau))$, which implies 
\begin{align}
V^t_{\beta,\lp}(\tau)=\beta^t\cl+ V^{t+1}_\beta(T(\tau))=\beta^t\cl+V^{t+1}_{\beta,\hp}(T(\tau))=V^t_{\beta,\hp}(\tau)
.
\end{align} In this case, 
\begin{align}
\label{eq:c2cplphp}
\cl+\beta V_{\beta,\hp}(T(\tau))=V_{\beta,\hp}(\tau).
\end{align}
Substitute \eqref{eq:transition} and \eqref{eq:hpevo} into \eqref{eq:c2cplphp}, we have
\begin{align}
\begin{split}
\label{eq:tau2}
& \tau=\frac{\cl-(1-\beta(1-\lambdana))(\rhn+\beta V_\beta(\lambdana))}{(1-(\lambdaaa-\lambdana)\beta)(\rha-\rhn+\beta( V_\beta(\lambdaaa)- V(\lambdana)))}\\&+\frac{\beta\lambdana(\rha+\beta V_\beta(\lambdaaa))}{{(1-(\lambdaaa-\lambdana)\beta)(\rha-\rhn+\beta( V_\beta(\lambdaaa)- V(\lambdana)))}}.
\end{split}
\end{align}

Next, we present how to compute $V_\beta(\lambdana)$ and $V_\beta(\lambdaaa)$.

Case 1: If $\lambdana\ge\tau$, then by Modeling Assumption~\ref{consumertransition}, $\lambdaaa\ge\lambdana\ge \tau$ and $p_F\ge\lambdana\ge\tau$. Thus, both $\lambdaaa$ and $\lambdana$ are in $\Phi_{\lp}$, therefore,
\begin{align}
\label{eq:v1}
V_\beta(\lambdana)=V_\beta(\lambdaaa)=\frac{\cl}{1-\beta}.
\end{align} 

Case 2:  If $\lambdana \le \tau$, we have $V_\beta(\lambdana)=V_{\beta,\hp}(\lambdana)$. Therefore,
\begin{align}
\label{eq:v21}
V_\beta(\lambdana)=(1-\lambdana)[\rhn+V^1_{\beta}(\lambdana)]+\lambdana[\rha+ V^1_{\beta}(\lambdaaa)].
\end{align} 
\begin{align}
& V_\beta(\lambdaaa)=\min\limits_{A_t\in\{\hp,\lp\}}{V_{\beta,A_t}(\lambdaaa)}\\
&=\min\{\cl+ V^1_{\beta}(T(\lambdaaa)),V_{\hp}(\lambdaaa)\}\\
&=\min\{\cl\frac{1-\beta^N}{1-\beta},\min\limits_{0\le n\le N-1}\{\cl\frac{1-\beta^n}{1-\beta}+V^{n}_{\beta, \hp}(T^{n}(\lambdaaa))\}\}.
\end{align}
Since $N\rightarrow \infty$ and $0\le \beta \le 1$,
\begin{align}
V_\beta(\lambdaaa)=\min\limits_{n>0}\{\cl\frac{1-\beta^n}{1-\beta}+\beta^n V_{\beta,\hp}(T^n(\lambdaaa))\}.
\end{align}
we have:
\begin{align}
\label{eq:v22}
& V_\beta(\lambdaaa)= \min\limits_{n\ge 0}\{\frac{\cl\frac{1-\beta^n}{1-\beta}+\beta^n[\bar{T}^n(\lambdaaa)(\rhn+C(\lambdana))+T^n(\lambdaaa)\rha]}{1-\beta^{n+1}[\bar{T}^n(\lambdaaa)\frac{\lambdana\beta}{1-(1-\lambdana)\beta}+T^n(\lambdaaa)]}\}.
\end{align}
where 
\begin{align}
T^n(\lambdaaa)=T(T^{n-1}(\lambdaaa)) = \frac{(\lambdaaa-\lambdana)^{n+1}(1-\lambdaaa)+\lambdana}{1-(\lambdaaa-\lambdana)},
\end{align}
\begin{align}
\bar{T}^n(\lambdaaa)=1-T^n(\lambdaaa)
\end{align}
\begin{align}
C(\lambdana)=\beta\frac{(1-\lambdana)\rhn+\lambdana \rha}{1-(1-\lambdana)\beta}.
\end{align}
\end{IEEEproof}

\section{Proof of Corollary~\ref{COR1}}
\label{PFCOR1}
\begin{IEEEproof}
By setting $V_{\lp}(p_t)\ge V_{\hp}(p_t)$, we have
\begin{equation}
\label{threshold}
\begin{split}
&\beta^t\cl+\beta V^t_\beta(T(p_t))\ge\\
&(1-p_t)[\beta^t\rhn+\beta V^t_{\beta}(\lambdana)]+p_t[\beta^t\rha+\beta V^t_{\beta}(\lambdaaa)].
\end{split}
\end{equation}
By Lemma~\ref{lm2} in the appendix, $V^t_\beta(p_t)$ is a concave function. Thus,  
\begin{equation}
\label{concavev}
\begin{split}
V^t_\beta(T(p_t))=V^t_\beta(\lambdana(1-p_t)+\lambdaaa p_t) & \\ \ge  (1-p_t)V^t_\beta(\lambdana)+p_t V^t_\beta(\lambdaaa).
\end{split}
\end{equation}
By substituting~\ref{concavev} into~\ref{threshold}, we can simplify inequality~\ref{threshold} to $(1-p_t)\rhn+p_t\rha\le \cl$, which implies $p_t\le \frac{\cl-\rhn}{\rha-\rhn}=\kappa$ when $V^t_{\lp}(p_t)\ge V^t_{\hp}(p_t)$.
\end{IEEEproof}

\section{Proof of Corollary \ref{CORBD}}
\label{PFCORBD}
\begin{IEEEproof}
Assume that $\lambdana\ge\tau$, we have $\lambdaaa> p_F=\frac{\lambdana}{1-(\lambdaaa-\lambdana)}>\lambdana\ge\tau$. In this case, By \eqref{eq:tau1} and \eqref{eq:v1}, we have 
\begin{align}
\label{eq:tau11}
\tau=\frac{\cl-\rhn}{\rha-\rhn}=\kappa.
\end{align}
Thus, $\tau=\kappa$ if $\lambdana>\kappa$.
Assume that $\lambdana<\tau$, then there are two cases for $p_F$:\\
Case 1: $p_F>\tau$, then $\lambdaaa>p_F>\tau$, which implies 
\begin{align}
\label{eq:v22lp}
V_\beta(\lambdaaa)=V_{\beta,\lp}(\lambdaaa)=\frac{\cl}{1-\beta}.
\end{align} 
By \eqref{eq:tau1}, \eqref{eq:v21}, and \eqref{eq:v22lp}, we have
\begin{align}
\label{eq:tau12}
\tau=\frac{\beta(\cl-\rha)\lambdana+\cl-\rhn}{(1-\beta)\rha-\rhn+\beta\cl}.
\end{align}
Therefore, $\tau=\frac{\beta(\cl-\rha)\lambdana+\cl-\rhn}{(1-\beta)\rha-\rhn+\beta\cl}$ if $p_F=\frac{\lambdana}{1-(\lambdaaa-\lambdana)}\ge\tau=\frac{\beta(\cl-\rha)\lambdana+\cl-\rhn}{(1-\beta)\rha-\rhn+\beta\cl}$ and $\lambdana<\frac{\beta(\cl-\rha)\lambdana+\cl-\rhn}{(1-\beta)\rha-\rhn+\beta\cl}$.
\\
Case 2: $p_F<\tau$, $\tau$ can be computed by \eqref{eq:tau2}, \eqref{eq:v21}, and \eqref{eq:v22}. Moreover, for fixed $\lambdaaa$, \eqref{eq:tau2} is a non-decreasing function w.r.t. $\lambdana$. Thus, let $\tau^+=\frac{\lambdana}{1-(\lambdaaa-\lambdana)}=\frac{\beta(\cl-\rha)\lambdana+\cl-\rhn}{(1-\beta)\rha-\rhn+\beta\cl}$,  $\tau\le\tau^+$ in Case 2. Therefore, $\tau^+$ is an upperbound for the optimal action in Case 2. 

Since \eqref{eq:tau2} is non-decreasing, \eqref{eq:tau12} is decreasing and intersects with \eqref{eq:tau11} at $\lambdana=\frac{\cl-\rhn}{\rha-\rhn}$, we have proved Corollary \ref{CORBD}.
\end{IEEEproof}
\section{Proof of Theorem \ref{THMCDT}}
\label{sec:app:cpdthm}
\begin{IEEEproof}
Let $p_F=\frac{\lambdana}{1-(\lambdaaa-\lambdana)}$ and $p'_F=\frac{\lambdapna}{1-(\lambdapaa-\lambdapna)}$ be the stationary belief of a \consumer\ being in alerted state when the transition matrix is $\mbf{\Lambda_{LP}}$ and $\mbf{\Lambda_{HP}}$. Since  $\tau$ be the threshold of offering either $\hp$ or $\lp$ coupons. Then we have $V^t_{\beta, \lp}(\tau)=V^t_{\beta, \hp}(\tau)$. This implies:
\begin{align}
\begin{split}
& V^t_{\beta, \lp}(\tau)-V^t_{\beta, \hp}(\tau)\\ &=\beta^t(\cl-(1-\tau)\rhn-\tau \rha)+[V^{t+1}_{\beta}(T(\tau))-V^{t+1}_{\beta}(T'(\tau))].
\end{split}
\end{align}
In order to compute the threshold $\tau$, we need to divide the computation into four cases with respect to to $T(\tau)$ and $T'(\tau)$.
%

Case 1: $T(\tau)>\tau$ and $T'(\tau)>\tau$. Thus
\begin{align}
V^{t+1}_{\beta}(T(\tau))=V^{t+1}_{\beta, \lp}(T(\tau))=\beta^{t+1}(\frac{\cl}{1-\beta}).
\end{align}
\begin{align}
V^{t+1}_{\beta}(T'(\tau))=V^{t+1}_{\beta, \lp}(T'(\tau))=\beta^{t+1}(\frac{\cl}{1-\beta}).
\end{align}
By setting $V^t_{\beta, \lp}(\tau)-V^t_{\beta, \hp}(\tau)=0$, we have $\tau=\frac{\cl-\rhn}{\rha-\rhn}$.

Case 2: $T(\tau)<\tau$ and $T'(\tau)>\tau$.

Since $T(\tau)<\tau$, $\hp$ coupons will be offered from timeslot $t+1$. Define $\eta=\frac{\beta(\lambdaaa-\lambdana)}{1-\beta(\lambdaaa-\lambdana)}$ and $\eta'=\frac{\beta(\lambdapaa-\lambdapna)}{1-\beta(\lambdapaa-\lambdapna)}$. Thus,
\begin{align}
& V^{t+1}_{\beta}(T(\tau))=V^{t+1}_{\beta,\hp}(T(\tau))\\&=\beta^t\sum\limits_{i=1}^\infty\{\beta^i[(\rha-\rhn)(p_F(1-(\lambdaaa-\lambdana)^i)+(\lambdaaa-\lambdana)^i\tau)+(\rhn)]\}\\
&=\beta^t\{\sum\limits_{i=1}^\infty\beta^i[(\rha-\rhn)(p_F+\rhn)]+\sum\limits_{i=1}^\infty\beta^i[(\rha-\rhn)(\tau-p_F)(\lambda
_1-\lambdana)^i]\}\\
&=\beta^t\{\frac{\beta}{1-\beta}[(\rha-\rhn)(p_F)+(\rhn)]\eta(\rha-\rhn)(\tau-p_F)\}\\
&=\beta^t\{p_F(\rha-\rhn)(\frac{\beta}{1-\beta}-\eta)\frac{\beta}{1-\beta}(\rhn)+\eta(\rha-\rhn)\tau\}.
\end{align}
Because $T'(\tau)>\tau$, only $\lp$ coupons will be offered after time $t$.
\begin{align}
V^{t+1}_{\beta}(T'(\tau))=V^{t+1}_{\beta,\lp}(T'(\tau))=\beta^t\frac{\beta \cl}{1-\beta}.
\end{align}
By setting $V^t_{\beta, \lp}(\tau)-V^t_{\beta, \hp}(\tau)=0$, we have then that $\tau$ is equal to 
\begin{align}
 \label{tau2}
\tau=\frac{(\cl-\rhn)+p_F(\rha-\rhn)[\frac{\beta}{1-\beta}-\eta]+\frac{\beta}{1-\beta}(\rhn-\cl)}{(\rha-\rhn)[1-\eta]}.
 \end{align}

Case 3: $T(\tau)<\tau$ and $T'(\tau)<\tau$.
In this case,
\begin{align}
V^{t+1}_{\beta}(T(\tau))=V^{t+1}_{\beta, \hp}(T(\tau)).
\end{align}
\begin{align}
V^{t+1}_{\beta}(T'(\tau))=V^{t+1}_{\beta, \hp}(T'(\tau)).
\end{align}
Setting $V^t_{\beta, \lp}(\tau)-V^t_{\beta, \hp}(\tau)=0$, we can find the threshold $\tau$ by equation 
 \begin{align}
 \label{tau3}
\tau=\frac{\cl-\rhn+(\rha-\rhn)[\frac{\beta}{1-\beta}(p_F-p'_F)-(p_F\eta-p'_F\eta')]}{(\rha-\rhn)[\eta'-\eta+1]}. 
 \end{align}


Case 4: $T(\tau)>\tau$ and $T'(\tau)<\tau$.
In this case,
\begin{align}
V^{t+1}_{\beta}(T(\tau))=V^{t+1}_{\beta, \lp}(T(\tau)).
\end{align}
\begin{align}
V^{t+1}_{\beta}(T'(\tau))=V^{t+1}_{\beta, \hp}(T'(\tau)).
\end{align}
By setting $V^t_{\beta, \lp}(\tau)-V^t_{\beta, \hp}(\tau)=0$, we have $\tau$ equals to equation 
 \begin{align}
  \label{tau4}
 \tau=\frac{(\cl-\rhn)(1+\frac{\beta}{1-\beta})-p'_F(\rha-\rhn)[\eta'-\frac{\beta}{1-\beta}]}{(\rha-\rhn)[1+\eta']}.
  \end{align}
\end{IEEEproof}
\bibliographystyle{IEEEtran}
\bibliography{IEEEabrv,Bibliography}
\end{document}